%% file: main_PRL_v3.0_arXiv.tex
\documentclass[aps,prl,preprint,superscriptaddress]{revtex4-2}

\usepackage{amsmath,amssymb,bm,mathrsfs}

\usepackage{graphicx}
\usepackage{dcolumn}      
\usepackage{multirow}
\usepackage{booktabs}
\usepackage{siunitx}
\usepackage{textcomp}
\usepackage{dblfloatfix}  
\usepackage[normalem]{ulem}
\usepackage{soul}
\usepackage{xcolor}

\usepackage{algorithm}
\usepackage{algpseudocode}

\usepackage{comment}
\usepackage{lipsum}
\usepackage{blindtext}
\usepackage[colorlinks=true,linkcolor=blue,citecolor=blue,urlcolor=blue]{hyperref}

\newcommand{\gag}{g_{a\gamma}}
\renewcommand{\arraystretch}{1.5}

\def \aaps {A\&AS}

\def \apj  {ApJ}
\def \apjs {ApJS}
\def \apjl {ApJL}

\def \nat  {Nature}

\begin{document}

\title{NuSTAR as an Axion Helioscope}

\author{J. Ruz}\email{Corresponding author: Jaime.Ruz@cern.ch}\affiliation{Fakult{\"{a}}t f{\"{u}}r Physik. Technische Universit{\"{a}}t  Dortmund. Dortmund, \textit{D-44221}, Germany}\affiliation{Centro de Astropart{\'i}culas y F{\'i}sica de Altas Energ{\'i}as. University of Zaragoza. Zaragoza, \textit{50009}, Spain}

\author{E. Todarello}\email{Corresponding author: elisa.todarello@nottingham.ac.uk}
\affiliation{School of Physics and Astronomy, University of Nottingham, University Park, NG7 2RD, Nottingham, United Kingdom}\affiliation{Dipartimento di Fisica. University of Torino, \textit{I-10125}, Torino, Italy}\affiliation{Istituto Nazionale di Fisica Nucleare. Sezione di Torino. \textit{I-10125}, Torino, Italy}

\author{J. K. Vogel}\email{Corresponding author: Julia.Vogel@cern.ch}
\affiliation{Fakult{\"{a}}t f{\"{u}}r Physik. Technische Universit{\"{a}}t  Dortmund. Dortmund, \textit{D-44221}, Germany}\affiliation{Centro de Astropart{\'i}culas y F{\'i}sica de Altas Energ{\'i}as. University of Zaragoza. Zaragoza, \textit{50009}, Spain}

\author{F. R. Cand\'on}
\affiliation{Fakult{\"{a}}t f{\"{u}}r Physik. Technische Universit{\"{a}}t  Dortmund. Dortmund, \textit{D-44221}, Germany}
\affiliation{Centro de Astropart{\'i}culas y F{\'i}sica de Altas Energ{\'i}as. University of Zaragoza. Zaragoza, \textit{50009}, Spain}

\author{M. Giannotti}
\affiliation{Centro de Astropart{\'i}culas y F{\'i}sica de Altas Energ{\'i}as. University of Zaragoza. Zaragoza, \textit{50009}, Spain}\affiliation{Department of Chemistry and Physics. Barry University. Miami Shores, \textit{33161}, Florida, USA}

\author{B. Grefenstette}
\affiliation{California Institute of Technology. Space Radiation Lab. Pasadena, \textit{91125}, California, USA}

\author{H. S. Hudson}
\affiliation{School of Physics and Astronomy. University of Glasgow. Glasgow, \textit{G12 8QQ}, Scotland, UK}

\author{I. G. Hannah}
\affiliation{School of Physics and Astronomy. University of Glasgow. Glasgow, \textit{G12 8QQ}, Scotland, UK}

\author{I. G. Irastorza}
\affiliation{Centro de Astropart{\'i}culas y F{\'i}sica de Altas Energ{\'i}as. University of Zaragoza. Zaragoza, \textit{50009}, Spain}

\author{C. S. Kim}
\affiliation{Department of Physics. University of California Santa Barbara. Santa Barbara, \textit{93106}, California, USA}

\author{M. Regis}
\affiliation{Dipartimento di Fisica. University of Torino, \textit{I-10125}, Torino, Italy}\affiliation{Istituto Nazionale di Fisica Nucleare. Sezione di Torino. \textit{I-10125}, Torino, Italy}

\author{D. M. Smith}
\affiliation{Physics Department and Santa Cruz Institute for Particle Physics. University of California Santa Cruz. Santa Cruz, \textit{95064}, California, USA}

\author{M. Taoso}
\affiliation{Istituto Nazionale di Fisica Nucleare. Sezione di Torino. \textit{I-10125}, Torino, Italy}

\author{J. Trujillo Bueno}
\affiliation{Instituto de Astrofísica de Canarias. La Laguna, \textit{38205}, Tenerife, Spain}\affiliation{Consejo Superior de Investigaciones Científicas, Spain}


\date{\today}


\begin{abstract}
We present a novel approach to investigating axions and axion-like particles (ALPs) by studying their potential conversion into X-rays within the Sun’s atmospheric magnetic field. Utilizing high-sensitivity data from the Nuclear Spectroscopic Telescope Array (NuSTAR) collected during the 2020 solar minimum, along with advanced solar atmospheric magnetic field models, we establish a new limit on the axion-photon coupling strength $g_{a\gamma}\lesssim 7.3\times 10^{-12}$~GeV$^{-1}$ at 95\% CL for axion masses $m_a\lesssim 4\times 10^{-7}$\,eV. This constraint surpasses current ground-based experimental limits, studying previously unexplored regions of the axion-photon coupling parameter space up to masses of $m_a\lesssim 3.4\times 10^{-4}$\,eV. These findings mark a significant advancement in our ability to probe axion properties and strengthen indirect searches for dark matter candidates. 
\end{abstract}

\maketitle
\textit{Introduction.---}
The Standard Model (SM) of particle physics has provided a remarkably successful framework for describing the fundamental particles and their interactions via three of the four known fundamental forces. Despite its predictive power, the SM leaves several key questions unresolved, notably the strong charge-parity (CP) problem in quantum chromodynamics (QCD) and the composition of dark matter. Intriguingly, a single theoretical extension proposed over four decades ago by Peccei and Quinn~\cite{1977PhRvL..38.1440P} offers a potential solution to both. This framework predicts the existence of a light, pseudo-scalar boson—the QCD axion~\cite{PhysRevLett.40.223, PhysRevLett.40.279, ABBOTT1983133, Preskill:1982cy,Dine:1981rt,Dine:1982ah}. Beyond this minimal realization, a broader class of particles known as axion-like particles (ALPs)~\cite{Jaeckel:2010ni} arises in many extensions of the SM. Like QCD axions, ALPs are pseudo-scalar bosons that couple to photons, but unlike QCD axions, they are not constrained by a specific relationship between their mass $m_a$ and the Peccei-Quinn symmetry-breaking scale $f_a$. Although ALPs do not solve the strong CP problem, they remain compelling candidates for dark matter. For the purposes of this Letter, we will refer to both QCD axions and ALPs generically as “axions.”

Numerous laboratory and astronomical experiments have been carried out to detect axions ~\cite{Sumico,PhysRevLett.133.221005, Asztalos_2010,Du_2018, Braine_2020, Bartram_2021, Boutan_2018, Bartram_2023, PhysRevD.42.1297, Zhong_2018, Backes_2021, haystaccollaboration2023new, Lee_2020, Jeong_2020, Kwon_2021, Lee_2022, Kim_2023, Yi_2023, Yang_2023, kim2023experimental, PhysRevX.14.031023, Quiskamp_2022, Abeln:2021, McAllister_2017, Alesini_2019, Alesini_2021, Di_Vora_2023, Devlin:2021fpq,Ouellet_2019, Salemi_2021, Crisosto:2019fcj}, with additional efforts either ongoing or proposed~\cite{RBahre_2013,PhysPotIAXO,Abeln:2021,Armengaud:2014gea}. These investigations predominantly focus on the interaction of axions with photons~\cite{Sikivie:1983ip}, described by the effective Lagrangian term $\mathcal{L}_{a\gamma} = g_{a\gamma}\, {\bf E}\cdot {\bf B}\, a$,
\begin{figure}[!t]
    \centering
    \includegraphics[width=0.7\textwidth]{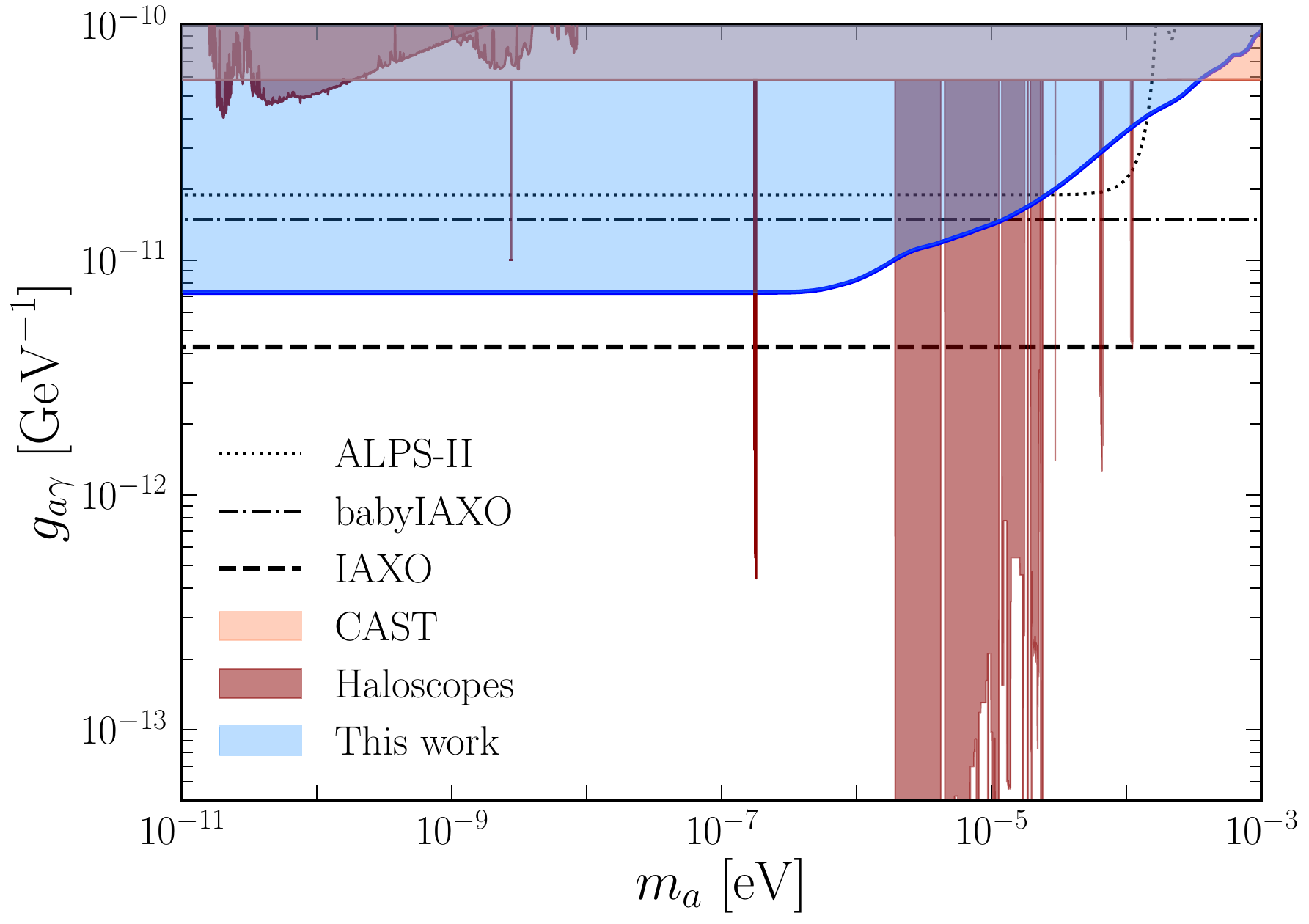}
    \caption{NuSTAR's 95\% CL exclusion on axion-photon coupling strength  $g_{a\gamma }$ (blue line). Regions excluded by this work are shown in shaded blue. We present our results in comparison to current~\cite{
    PhysRevLett.133.221005, Asztalos_2010,Du_2018, Braine_2020, Bartram_2021, Boutan_2018, Bartram_2023, PhysRevD.42.1297, Zhong_2018, Backes_2021, haystaccollaboration2023new, Lee_2020, Jeong_2020, Kwon_2021, Lee_2022, Kim_2023, Yi_2023, Yang_2023, kim2023experimental, PhysRevX.14.031023, Quiskamp_2022, Abeln:2021, McAllister_2017, Alesini_2019, Alesini_2021, Di_Vora_2023, Devlin:2021fpq,Ouellet_2019, Salemi_2021, Crisosto:2019fcj} and future~\cite{RBahre_2013,PhysPotIAXO,Abeln:2021,Armengaud:2014gea} laboratory experiments.}\label{fig:exclusion_main}
\end{figure}
 where $a$ represents the axion field, ${\bf E}$ and $\bf{B}$ denote electric and magnetic fields, respectively, and $g_{a\gamma}$ is the coupling strength. Such an interaction enables the Primakoff effect, where photons convert into axions (and vice versa) in the presence of an electromagnetic field. 
 
The Sun acts as a bright source of axions, which are generated in its core via the Primakoff process, wherein thermal photons are converted into axions through interactions with the electric fields of electrons and ions in the plasma. These axions can then reconvert into photons within the Sun's atmospheric magnetic fields, yielding a distinct X-ray flux with spectral and spatial characteristics that allow differentiation from typical solar X-ray emissions and background signals in X-ray telescopes. Advances in solar X-ray observational capabilities, such as higher resolution and sensitivity, have improved prospects for identifying solar-origin axions. Initial efforts using satellite instruments for axion detection include the \textit{Yohkoh}/SXT ~\cite{1991SoPh..136...37T} mission data from 1991-2001, utilized in~\cite{Carlson:1995xf} to establish early constraints on the axion-photon coupling. Later missions, including RHESSI~\cite{2007ApJ...659L..77H,2010ApJ...724..487H} and \textit{Hinode}/XRT \cite{2012ASPC..455...25H}, applied similar methodologies, gradually refining observational techniques and constraints on axion interactions. However, due to the high background rates and optimization of these instruments for bright solar X-rays, dedicated solar missions are not optimized for weak-signal dark matter searches.\par
In this letter, we present the first analysis of solar observation X-ray data from NASA's Nuclear Spectroscopic Telescope Array (NuSTAR)~\cite{2013ApJ...770..103H}, specifically aimed at searching for signatures of solar axion conversion in the solar atmosphere. We find no evidence for an axion-induced signal above the expected background, allowing us to place stringent constraints on the axion-photon coupling over a significant region of parameter space. In particular, we constrain the axion-photon coupling $g_{a\gamma}$ to less than  
$7.3\times 10^{-12}$~GeV$^{-1}$ at the $95\%$ confidence level (CL) for axion masses $m_a \lesssim 4\times 10^{-7}$~eV, substantially improving upon previous limits set by laboratory-based experiments~\cite{Sumico,PhysRevLett.133.221005, Asztalos_2010,Du_2018, Braine_2020, Bartram_2021, Boutan_2018, Bartram_2023, PhysRevD.42.1297, Zhong_2018, Backes_2021, haystaccollaboration2023new, Lee_2020, Jeong_2020, Kwon_2021, Lee_2022, Kim_2023, Yi_2023, Yang_2023, kim2023experimental, PhysRevX.14.031023, Quiskamp_2022, Abeln:2021, McAllister_2017, Alesini_2019, Alesini_2021, Di_Vora_2023, Devlin:2021fpq,Ouellet_2019, Salemi_2021, Crisosto:2019fcj}, exceeding the results of planned experiments~\cite{RBahre_2013,PhysPotIAXO,Abeln:2021,Armengaud:2014gea}
and competing with current astrophysical limits ~\cite{Li:2024zst, Li_2022, Reynolds_2020, Abramowski_2013, Ajello_2016, Davies_2023, Jacobsen:2022swa,Li:2024zst,Xiao:2020pra,Ning:2024ek,Dessert_2022_1,Dessert_2022_2,Noordhuis:2022ljw,Dessert_2020,Hoof_2023,Manzari:2024jns,Foster:2022fxn, Battye:2023oac,Escudero:2023vgv, Fox:2023xgx}.
Fig.~\ref{fig:exclusion_main} shows the outcome of the present study in comparison to those from previous and forthcoming ground-based axion experiments. 

\vspace{0.1cm}
\textit{Axion Conversion in Solar Atmospheric Magnetic Field---}
Accurate theoretical models for the solar axion flux are well-established, and our analysis is grounded in the results presented in~\cite{Raffelt2008, Carenza:2024ehj}. Since both, the generation of axions from X-rays in the solar core and their reconversion into photons in the solar atmosphere's magnetic field occur via the Primakoff process, the resulting X-ray signal detected by satellite missions scales with $\gag^{4}$. Axions generated in the core are highly relativistic and follow approximately radial trajectories to the solar surface. The probability of their conversion into photons is then provided by~\cite{Raffelt:1987im},
\begin{equation}
\label{eq:probMainnew}
P_{a\rightarrow\gamma}(E,h,m_{a}) = \frac{1}{4}\gag^2\Big|\int_0^h dh' B_\perp(h')\ e^{i\int_0^{h'}dh''q(h'',m_{a})}\ e^{-\frac{1}{2}\int_{h'}^h dh''\Gamma(h'')}
\Big|^2 ,\enspace 
\end{equation}
where $h$ is the altitude above the visible surface of the Sun,  $B_\perp$ is the component of the magnetic field perpendicular to the direction of propagation of the photon, and $q(h,m_{a}) = (\omega_{p}^2(h) - m_a^2)/2E$ is the momentum exchanged between the photon in the medium and the axion, both carrying energy $E$. The plasma frequency of the medium is
$\omega_{p}(h) = \sqrt{e^2 n_e(h)/m_e}$, where $e$ and $m_e$ are the electric charge and mass of the electron, respectively, while $n_e(h)$ is the height-dependent number density of free electrons plus that of bound electrons with ionization energy much smaller than the X-ray energy. Notably, in natural units, the plasma frequency corresponds to an effective photon mass.  
The term containing $\Gamma$ represents the absorption between altitudes $h'$ and $h$ of the arising X-rays in the solar atmosphere~\cite{Raffelt:1987im, PhysRevD.39.2089}. 
 
\begin{figure}[!t]
\centering
\begin{minipage}{.6\textwidth}
  \centering
  \hspace{-0.5cm}
  \includegraphics[width=1.\linewidth]{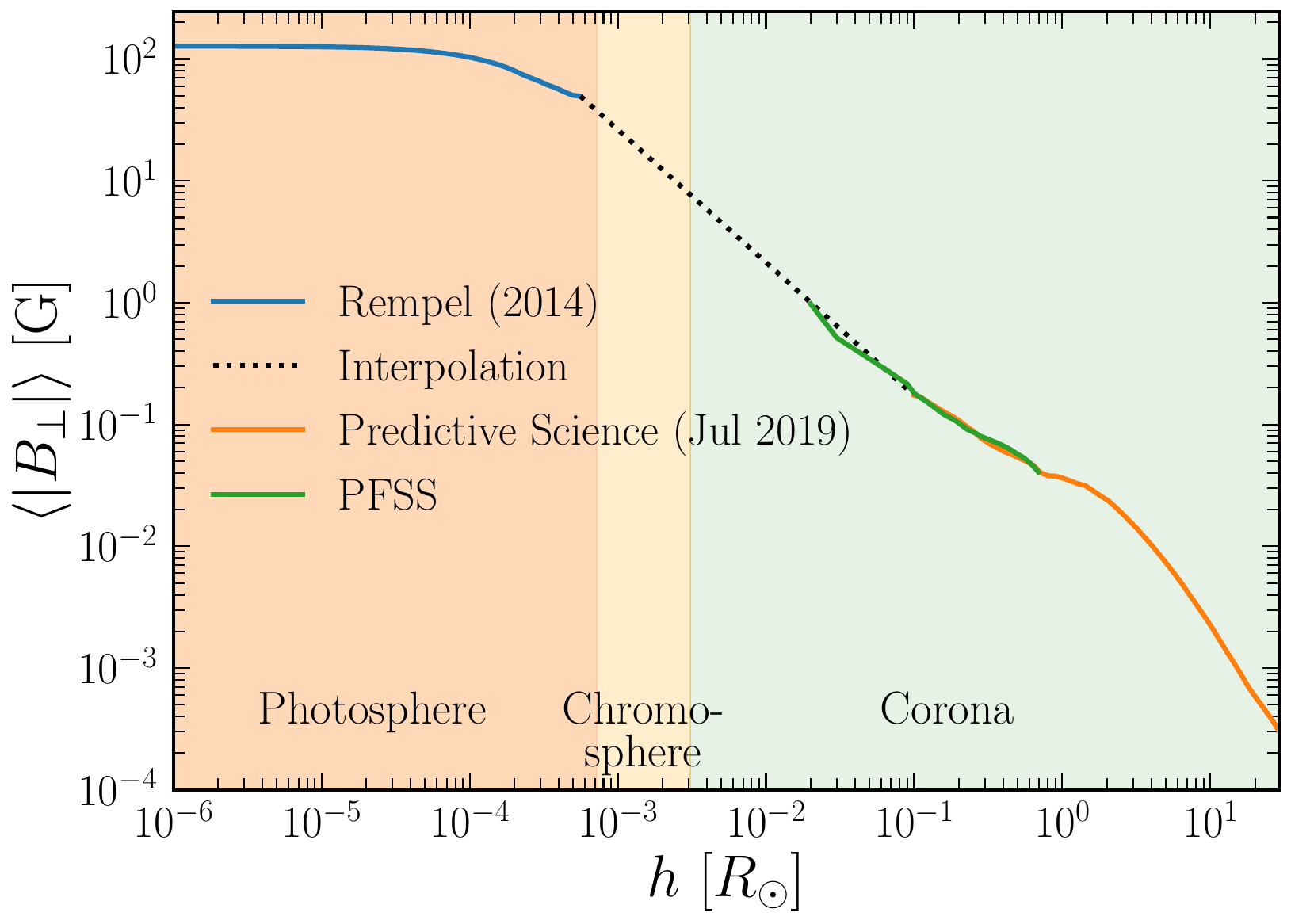}
\end{minipage}
\\
\vspace{0.25cm}
\begin{minipage}{.6\textwidth}
  \centering
    \hspace{-0.5cm}
  \includegraphics[width=1.\linewidth]{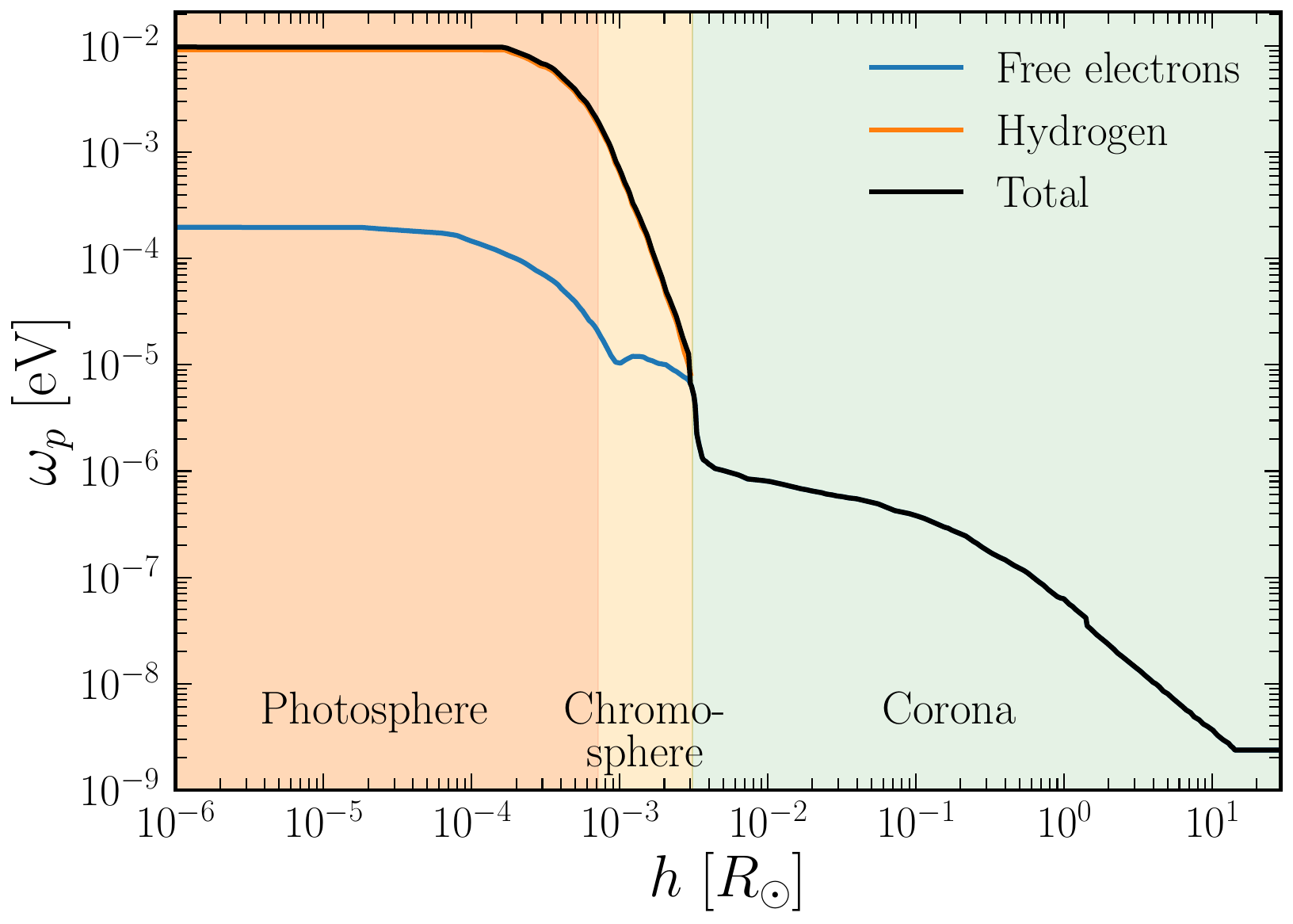}
\end{minipage}
\caption{Modeling of the transverse component of the solar atmospheric magnetic field (\textbf{top})~\cite{Rempel_2014,ps2017}, and contributions to axion plasma frequency (\textbf{bottom}) for the photosphere (orange), chromosphere (yellow), and corona (green) regions as function of height in units of the solar radius.}
\label{fig:models}
\end{figure}

In this work, we characterize the solar magnetic field by means of advanced and realistic Magneto-Hydrodynamics (MHD) simulations that accurately represent the quiet-Sun atmosphere \textemdash that corresponds to the conditions for the NuSTAR solar observations used in this letter. 
For altitudes below 400 km, we make use of magneto-convection simulations with small-scale dynamo activity~\cite{Rempel_2014} developed with the MURaM radiative MHD code~\cite{2005A&A...429..335V, 2009ApJ...691..640R}. This three-dimensional (3D) model applies periodic boundary conditions in the horizontal plane orthogonal to the solar radius vector at the observed solar disk location. It accurately represents the quiet inter-network regions of the solar photosphere, which cover most of the solar disk across the solar activity cycle. The model's magnetic field, tangled at scales beneath current telescope resolution, shows a mean field strength of 170 Gauss at the visible surface, matching observed depolarization effects in photospheric spectral line polarization~\cite{2004Natur.430..326T, PinoAleman2018}. The blue curve in the top panel of Fig.~\ref{fig:models} illustrates the altitude variation of $\langle \lvert B_{\perp} \rvert \rangle$, calculated as the spatial average of $\lvert B_{\perp} \rvert$ across all points in the plane perpendicular to the solar radius vector.
For modeling the coronal magnetic field, we utilize the publicly available Predictive Science Inc. (PSI) MHD simulation corresponding to the solar eclipse of July 2, 2019~\cite{ps2017}, and apply it to  the quiet phase of solar activity of our observations. See Supplemental Material (SM) for further details~\cite{supplemental_jra}.
The PSI simulation provides a 3D representation of the coronal magnetic field up to an altitude of $h=29 R_\odot$. To obtain $\langle|B_\perp|\rangle$, we extract the perpendicular component of the magnetic field relative to the Earth-Sun line of sight, averaged over a disk of radius $0.1R_\odot$ centered on our signal region, as represented by the orange line in the top panel of Fig.~\ref{fig:models}. We perform a power-law interpolation between the photospheric and coronal magnetic field models (dashed line in the top panel of Fig.~\ref{fig:models}), which is further validated by the Potential-Field-Source-Surface (PFSS) model for the specific day of our observation~\cite{pffs6}, denoted by the green line in the top panel of Fig.~\ref{fig:models}.

To accurately calculate the axion-photon conversion probability in the solar magnetic atmosphere, it is essential to model the free electron and hydrogen density. For this purpose, we utilize the model from~\cite{2008GeofI..47..197D}, which is representative of the quiet Sun. This model builds on the VAL-C chromospheric model~\cite{1981ApJS...45..635V} and the coronal model~\cite{1976RSPTA.281..339G}. Our focus is specifically on the quiet Sun, excluding the added coronal density contributions from the streamer belt. In addition to free electrons and hydrogen, we account for helium (He) in the plasma density, assuming it follows the density profile of hydrogen (H) with a relative abundance $n_{\rm{H}} / n_{\rm{He}} = 0.06$~\cite{2012ApJ...755...33S, 2011SoPh..268..255C, 2009LanB...4B..712L}. Contributions from heavier elements are not included in the computation of $q$, as their low abundances render their impact on the plasma frequency negligible. In the chromosphere, where the ionization fraction is low, the plasma frequency is primarily determined by the density of bound electrons in atoms, mainly hydrogen, as shown in the bottom panel of Fig.~\ref{fig:models}. In the corona, with temperatures on the order of $10^{6}$~K, hydrogen and helium are fully ionized. Given the higher mass of nuclei, their contribution to the plasma frequency is minimal compared to that of free electrons.

The absorption coefficient $\Gamma$ is calculated as the sum over all species $i$ such that $\Gamma_i = n_i\sigma_i$, where $n_i$ represents the number density of the species $i$ and $\sigma_i$ is the total photon cross-section for that species. We include elements up to atomic number $Z=30$ in our calculation, using elemental abundances from the CHIANTI database ~\cite{1997A&AS..125..149D, 2021ApJ...909...38D}, specifically based on the Schmelz extended model of coronal abundances derived in~\cite{2012ApJ...755...33S, 2011SoPh..268..255C, 2009LanB...4B..712L}. For each element, we obtain the total photon cross-section from the NIST XCOM database~\cite{xcom}, which accounts for X-ray attenuation due to coherent (Rayleigh) and incoherent (Compton) scattering, as well as photoelectric absorption.

\vspace{0.1cm}
\textit{NuSTAR observations.---}\label{sec:Nobservations}
On February 21, 2020, NuSTAR was directed toward the center of the solar disk, capturing 23,814 seconds of live exposure data with flight module $A$ and 24,983 seconds with flight module $B$, to support both a solar axion search and the analysis of X-ray bright points on the solar photosphere~\cite{sarah24}. The minimum of the 11-year solar activity cycle provided a unique opportunity to study the emission from the quiet Sun, and this campaign offers the best data for axion studies, since it involved a long dwell at the disk center during a period of very low solar activity, providing an ideal environment by minimizing interference from bright active regions or solar flares within the $10'\times10'$ field of view~\cite{2013ApJ...770..103H}. To identify the predicted axion spectrum, counts were accumulated from pixels within $0.1$ solar radii ($1.6'$) of the disk center, while background counts were gathered from an annulus between $0.15$ and $0.30$ solar radii. 

Although the solar axion surface luminosity profile~\cite{Andriamonje_2007} indicates that an optimal source region for our analysis, given NuSTAR's energy threshold, would be at approximately $0.15\,R_{\odot}$, we adopt a more conservative selection radius of $0.1\,R_{\odot}$ and account for remaining expected axion signal in the background region as discussed later on in the Analysis Section. This choice satisfies two key criteria: it maximizes the axion flux originating from the solar core while minimizing contamination from bright solar features present during the observation period.

\begin{figure}[t]
    \hspace{-0.75cm}
    \includegraphics[width=0.7\textwidth]{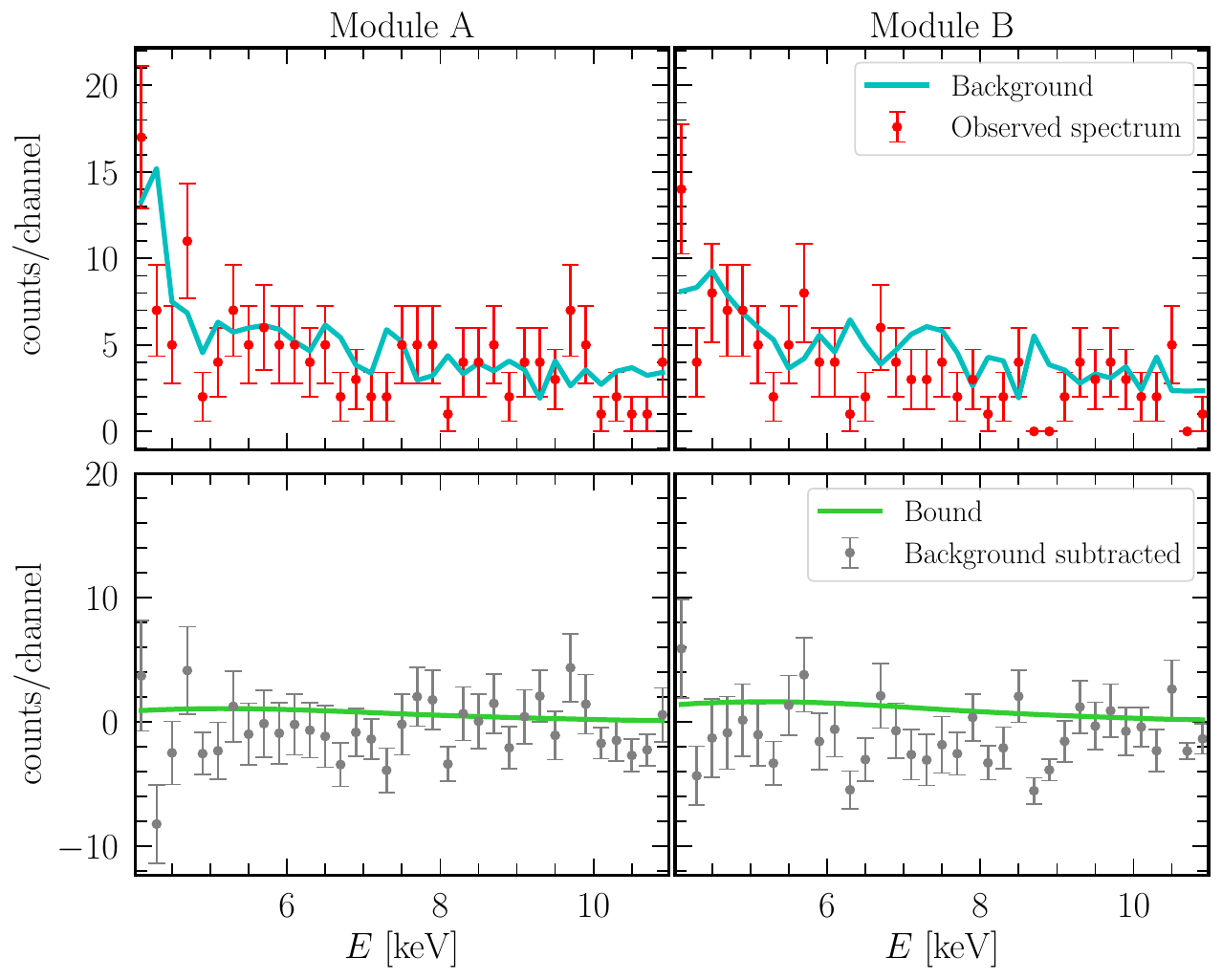}
   
    \caption{NuSTAR's observed spectra for module A (\textbf{left})
    and B(\textbf{right}) in the signal (red) and area normalized background (cyan) regions.
    Background-subtracted total spectrum (\textbf{bottom row}) and 95\% CL spectral shape of the axion component folded through the instrument response for an exemplary axion mass of $m_a = 10^{-7}$~eV is shown. %
    }
    \label{fig:spectroscopy}
\end{figure}

For the analysis presented here, NuSTAR conducted a total of nine observations (orbits). Modules $A$ and $B$ have been analyzed independently, since time exposure, effective area and detector response are inherent to each particular module. To account for the motion of the solar center in the field of view (FoV), each observation has been divided into four segments, each approximately 700 seconds in duration.
The top panels of Fig.~\ref{fig:spectroscopy} present the total counts in the signal region as a function of energy (red dots), along with the background (cyan line) derived from the outer annular region for module $A$ (left) and module $B$ (right).  The background of the respective modules was adjusted by normalizing to the source area, smoothing via a running polynomial fit, and further correcting for the known gradient of cosmic background X-rays across the detector chips. For clarity, statistical uncertainties of the plotted background were omitted \textemdash the detector background was measured over an area 5.3 times larger than the signal region, enhancing the accuracy of background estimation. The bottom panel shows the background subtracted spectrum (gray dots) including error propagation for modules $A$ (left) and $B$ (right) and we report, for illustrative purposes, the expected axion flux (green line) corresponding to the 95\% CL of $\gag$ for an axion with mass of $10^{-7}$~eV. Detailed procedures for the NuSTAR solar data processing using \texttt{NASA-HEASARC HEASoft 6.34}, NuSTAR Data Analysis Software (\texttt{NuSTARDAS v.2.1.4}) and calibration database (\texttt{CALDB v.20240325}) are discussed in SM \cite{supplemental_jra}. \texttt{XSPEC} tool \cite{arnaud1999xspec} version 12.13.1 was used to read the spectra and export the data into ASCII format.
 
\vspace{0.1 cm}
\textit{Data Analysis.---}\label{sec:methods}
The expected photon count resulting from axion conversion within a given NuSTAR energy bin $i$, originating from region $j=s,b$ \textemdash where $s$ denotes the signal region ($r<0.1R_\odot$) and $b$ the background region ($0.15R_\odot<r<0.3R_\odot$)\textemdash is computed separately for each module $k=A,B$ and observation segment as:

\begin{equation}
     N_{\gamma,i}^{j,k}(h,m_{a}) = \Delta  t^{k} \int dE\, A_{\rm{eff}}^{j,k}(E)\epsilon_{RMF,i}^{k}\,(E)\frac{dN_a^j}{dA\, dt\, dE}(E)\,P_{a\to\gamma}(E,h,m_{a})\enspace, 
\end{equation}

where $\Delta t^{k}$ is the observation exposure and $A_{\rm{eff}}^{j,k}$ the effective area of a given module $k$. The energy response function of each telescope in an energy bin $i$ is denoted by $\epsilon_{RMF,i}^{{k}}$ ~\cite{SPIE_NuSTAR_Effective}. The term $\frac{dN_a^j}{dA\, dt\, dE}(E)$ corresponds to the differential axion flux originating from the solar core, while $P_{a\to\gamma}(E,h,m_{a})$ represents the axion-photon conversion probability within the solar atmosphere following Eq.~\ref{eq:probMainnew}. We calculate the total number of photons from axion conversion expected in each module by summing over all the relative segments and account for leftover signal in $0.15-0.3R_\odot$. The likelihood analysis was performed using an 80 eV bin width. However, for improved visual clarity, Fig. \ref{fig:spectroscopy}  is presented with a bin width of 200 eV.

The total number of X-ray photons expected in the signal region for a given module and energy bin, is given by
\begin{equation}
\lambda^{k}_{i} = N_{\gamma,i}^{s,k} (g_{10}^{4})+ z^{s,k}_{i}\enspace,
\label{eq:nevents}
\end{equation}

where $z^{s,k}_{i}$ represents the background of bin $i$ in the signal region and $g_{10}\equiv g_{a\gamma}/10^{-10}{\,\rm GeV}^{-1}$.
To estimate $z^{s,k}_{i}$, we use the photon counts from the background region. Assuming that the background X-ray photon density is uniform across the solar disk, we can express this as 
\begin{equation}
z^{s,k}_{i}  
= (t^{k}_{i}- N_{\gamma,i}^{b,k}) \frac{A_\odot^s} {A_\odot^b} , 
\end{equation}

where $t^{k}_{i}$ denotes the total number of photons detected in the bin $i$ of the background region, $N_{\gamma,i}^{b,k}$ is the faint contribution from axions within the background annulus. Furthermore, $A_\odot^s$ and $A_\odot^b$ represent the areas of the signal and background regions, respectively.

 We assume independent Poisson statistics  for each energy bin and calculate the likelihood for each module as
\begin{equation}
\label{eq:PoisLike}
\mathcal{L}^{k} \propto \prod_i \frac{e^{-\lambda_i^{k}} \lambda_i^{n_i^{k}}}{n_i^k!}\enspace.
\end{equation}
In this analysis, $n_i^k$ denotes the number of photons detected in the $i$-bin of the energy spectrum of a given module. To derive our limit, we focus on the energy range from $4$ to $11$\,keV and apply the prior $\Theta(g_{10}^4)$, where $\Theta$ is the Heaviside step function. We then integrate the Bayesian posterior probability density function (PDF) from zero to the point that includes 95\% of the total PDF area. 
To obtain the combined bound shown in Figure~\ref{fig:exclusion_main}, we multiply the likelihoods of each module and follow the procedure described above applied to the total likelihood. The two modules yield similar bounds, the one from module $A$ being slightly more stringent. 

Our study considers various sources of systematic error, including the model uncertainties of the solar atmospheric magnetic field, which are the dominant contributors to the systematics, as well as uncertainties in the solar axion flux and the background of the NuSTAR detector.  
Overall, the combined effect of these uncertainties in our bound is calculated  to be $_{-21.2\%}^{+27.6\%}$, which is further discussed in SM \cite{supplemental_jra}.

\vspace{0.1cm}
\textit{Results and Discussion.---}
Using the statistical analysis outlined above, the NuSTAR solar axion observation data provides a constraint of $g_{a\gamma}\lesssim 7.3\times 10^{-12}$~GeV$^{-1}$ at 95\% CL on the axion-photon coupling strength $g_{a\gamma}$ for axion masses $m_a \lesssim 4\times 10^{-7}$~eV. This limit is depicted in blue in the broader $m_{a}-g_{a\gamma}$ parameter space shown in Fig.~\ref{fig:exclusion_main}. While NuSTAR’s limits remain nearly constant for $m_a \lesssim 4\times 10^{-7}$~eV, they weaken for higher axion masses due to the loss of coherence in the conversion probability, as described by Eq.~\eqref{eq:probMainnew}, when the axion mass exceeds the plasma frequency at lower solar atmospheric heights. At low axion masses, the $\rm{NuSTAR}$ limit represents a significant improvement over the leading CAST constraint~\cite{PhysRevLett.133.221005} (light red) and falls between the projected limits of future experiments, such as $\rm{BabyIAXO}$~\cite{Abeln:2021} (dashed-dotted, with data collection scheduled to begin in 2028) and IAXO~\cite{PhysPotIAXO} (dashed, expected in the mid 2030s). For comparison, we also report the projected sensitivity of the ALPS-II laser propagation experiment at DESY~\cite{RBahre_2013} (dotted line) in Fig.\ref{fig:exclusion_main}. We note that the NuSTAR analysis presented here explores the $m_{a}-g_{a\gamma}$ parameter space with unprecedented sensitivity over a broad range of axion masses. The region explored in this project includes areas not accessible to any other current or near future axion experiments and probes large regions that could address the transparency hints~\cite{transparency}.

Unlike haloscope experiments~\cite{Asztalos_2010,Du_2018, Braine_2020, Bartram_2021, Boutan_2018, Bartram_2023, PhysRevD.42.1297, Zhong_2018, Backes_2021, haystaccollaboration2023new, Lee_2020, Jeong_2020, Kwon_2021, Lee_2022, Kim_2023, Yi_2023, Yang_2023, kim2023experimental, PhysRevX.14.031023, Quiskamp_2022, Abeln:2021, McAllister_2017, Alesini_2019, Alesini_2021, Di_Vora_2023, Devlin:2021fpq,Ouellet_2019, Salemi_2021, Crisosto:2019fcj}, helioscopes, including our study, do not assume axions to be dark matter, as the axion flux is inherently produced by the Sun. To maintain consistency, the luminosity associated with the axion flux must remain a small fraction of the total solar luminosity, to avoid significant energy loss due to escaping axions~\cite{SCHLATTL1999353}. The measured solar neutrino flux and helioseismology observations place a constraint of $g_{10} \lesssim 4.1$~\cite{Vinyoles_2015}, which is well above the bounds reported here. In this regime, the theoretical precision of the axion flux is at the few-percent level~\cite{Hoof:2021mld}, as outlined above, and therefore does not pose a significant systematic concern.

\vspace{0.1cm}
\textit{Acknowledgments.---}
We gratefully acknowledge the support of the NuSTAR operations, software, and calibration teams in the execution and analysis of these observations. Our thanks also extend to Georg G. Raffelt, Ji{\v r}{\' i} {\v S}t{\v e}pan and T. O'Shea for insightful discussions. ET, MR, and MT are supported by the project ``Theoretical Astroparticle Physics (TAsP)" funded by INFN and the ``Grant for Internationalization" from the University of Torino. MT further acknowledges the research grant ``Addressing Systematic Uncertainties in Searches for Dark Matter No. 2022F2843L" funded by MIUR. ET expresses gratitude to the University of Zaragoza for its hospitality during the initial phases of this work. 
This publication is based on work supported by COST Action COSMIC WISPers CA21106.

The code to generate the results of this study is openly available at \cite{elisa_git}.
\quad
\bibliographystyle{apsrev4-2}
\bibliography{sn-bibliography_PRL_short}


\clearpage

\include{supplemental_PRL_v3.0_arXiv.tex}

\end{document}

%% file: supplemental_PRL_v3.0_arXiv.tex
\setcounter{page}{1}
\setcounter{figure}{0}
\setcounter{equation}{0}
\setcounter{table}{0}
\renewcommand{\theequation}{S\arabic{equation}}
\renewcommand{\thepage}{S\arabic{page}}
\renewcommand\thefigure{\Alph{figure}}

\clearpage 
\onecolumngrid 
\begin{center}
    \large \textbf{Supplemental material}
\end{center}
\vspace{0.6 cm}

\section{Modeling of the Axion Flux and Solar Atmosphere.} 
\subsection{Solar axion flux}\label{sec:solar_axion}

\begin{figure}[t!]
   \centering
    \includegraphics[width=0.6\textwidth]{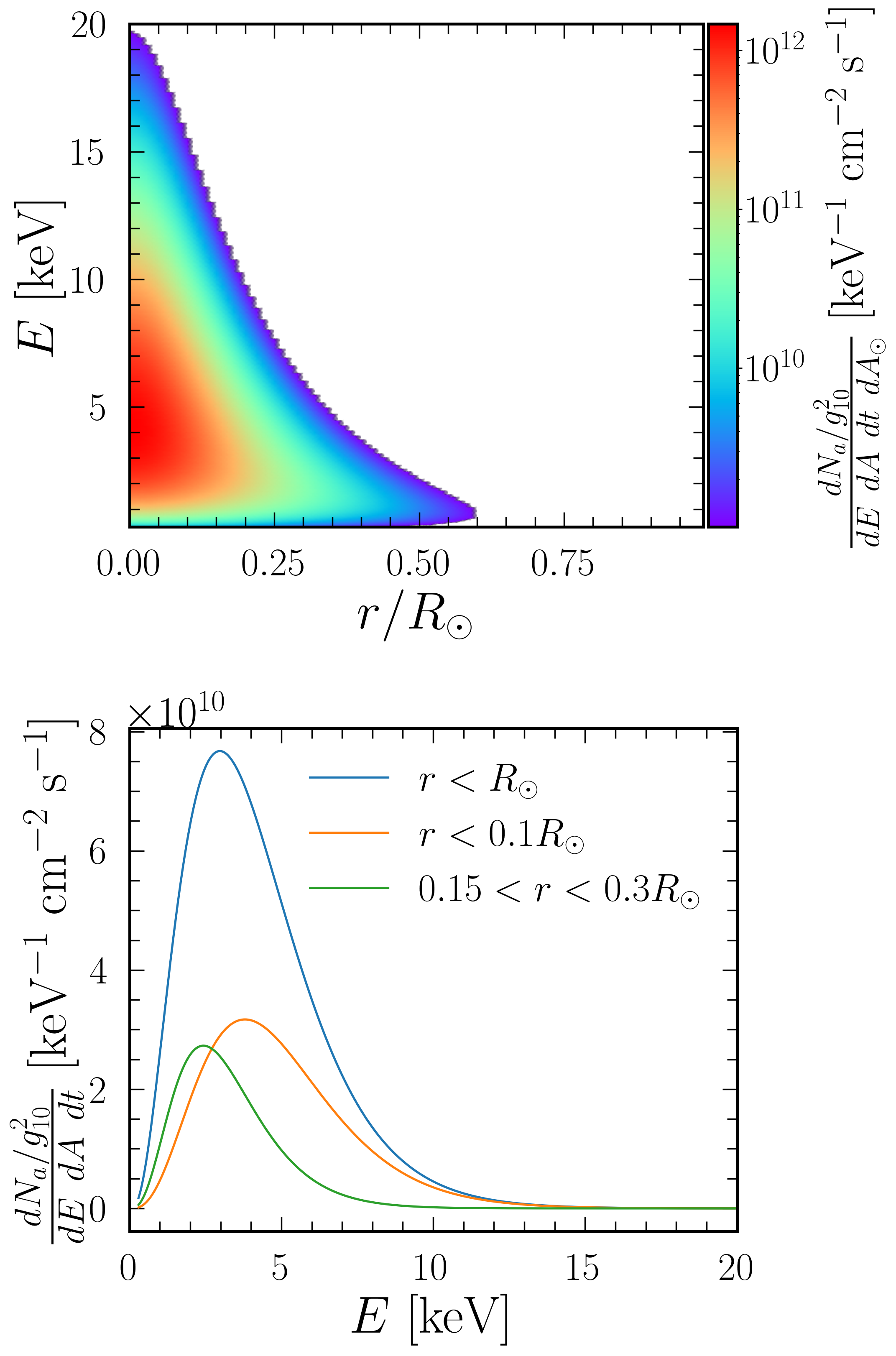}
    \caption{\textbf{Top:} Solar axion surface luminosity depending on energy and the radius $r/R_{\odot}$ on the solar disk. The flux ($\frac{dN_{a}/g^{2}_{10}}{dE\, dA\,dt\,dA_\odot}$) is given in units of axions $\rm{keV^{-1}cm^{-2}s^{-1}}$ per unit surface area on the solar disk. \textbf{Bottom:} Differential solar axion spectrum, derived by integrating the model shown on the top up to different values of $r/R_{\odot}$ in units of the solar radius $R_{\odot}$.}
    \label{fig:flux_plot}
\end{figure}

In the core of the Sun, black-body photons can convert to axions in the presence of the electric fields generated by charged particles in the high-temperature plasma. This process follows the interaction $\gamma+(e^{-}, Ze)\rightarrow a + (e^{-}, Ze)$~\cite{Sikivie:1983ip}. According to current models of stellar nucleosynthesis~\cite{bahcall1982standard}, nuclear fusion reactions within the Sun’s core serve as the primary energy source, with the proton-proton chain being a key mechanism. The core, characterized by a temperature of $T_{c}\sim 1.3\,\rm{keV}$ and a density of $\rho_{c}\sim 1.5 \times 10^{5}\,\rm{kg/m^{3}}$, occupies about 20\% of the solar radius and is the only region where significant heat production occurs. Axion emission from the solar core is highly temperature-dependent, with an average energy of $E_a\sim 2.7 \times T_c$, aligning with the temperature dependence of thermal X-ray production. Given the temperature gradient within the solar interior, the axion spectrum exhibits radial variation, as illustrated in Fig.~\ref{fig:flux_plot}. This model allows calculation of the total differential axion flux from the core, incorporating the standard solar model and the Sun’s luminosity $L_{\odot}=3.85\times 10^{33}\,\rm{erg\,s^{-1}}$ \cite{Raffelt2008, cast_jcap2009}.

The differential solar X-ray flux resulting from axion-photon conversion in the Sun’s atmosphere  depends on two key factors: the axion flux generated within the solar interior, and the probability of axions converting into photons upon passing through the solar atmosphere, denoted as $(P_{a\rightarrow\gamma})$, and can be written as:
\begin{equation}\label{eq:flux}
\frac{dN_{\gamma}}{dE\,dA\,dt\,d\Omega}=\frac{dN_{a}}{dE\,dA\,dt\,d\Omega}\,P_{a\rightarrow\gamma}\enspace,
\end{equation}
{with $dE,$ $dA$, $dt$, and $d\Omega$ being the elements of energy, area, time, and solid angle, respectively. As shown in Fig. \ref{fig:flux_plot}, the resulting axion flux peaks at energies $\sim 4{\, \rm keV}$ in the solar core. These features are imprinted on the spectral and morphological properties of the resulting X-ray emission in Eq.~\eqref{eq:flux}. The axion flux from Primakoff conversion in the Sun's core~\cite{cast_jcap2009} is shown on the top panel of Fig.~\ref{fig:flux_plot}, where $A_\odot$ represents the transverse area of the solar disk expressed in units of solar radius squared $R_\odot^2$. The bottom panel of Fig.~\ref{fig:flux_plot} shows the total axion flux and the axion flux arising from our signal and background regions.

\subsection{Solar magnetic field}\label{sec:results_corona}

To accurately model axion conversion in the solar atmosphere, we require detailed knowledge of the magnetic field structure.

As explained in the main manuscript, the magnetic field in the quiet solar photosphere is obtained from~\cite{Rempel_2014}. Among the various magneto-convection 3D models, we selected the one with a mean field strength of 170 G at the model’s visible surface. This selection was made because it is the only model consistent with observational constraints on scattering polarization signals in the Sr I 460.7 nm line~\cite{ps2017}, see~\cite{PinoAleman2018}.

The coronal magnetic field can be described by magnetohydrodynamic (MHD) models based on photospheric magnetic field observations.
We used the Predictive Science Inc. (PSI) model for the magnetic structure during the July 2, 2019, total solar eclipse, as detailed in~\cite{ps2017}. Although this model represents an epoch several months before the NuSTAR observations on February 21, 2020, it provides a relevant baseline, despite some solar evolution over that period (see the decrease in solar activity evident in the two snapshots from the Hinode X-Ray Telescope shown in Fig.\ref{fig:corona_pic}).  During the eclipse, an isolated active region was Earth-facing and represents the primary activity captured in the three-dimensional PSI model under consideration.

\begin{figure}[t!]
\centering
    \includegraphics[width=0.6\textwidth]{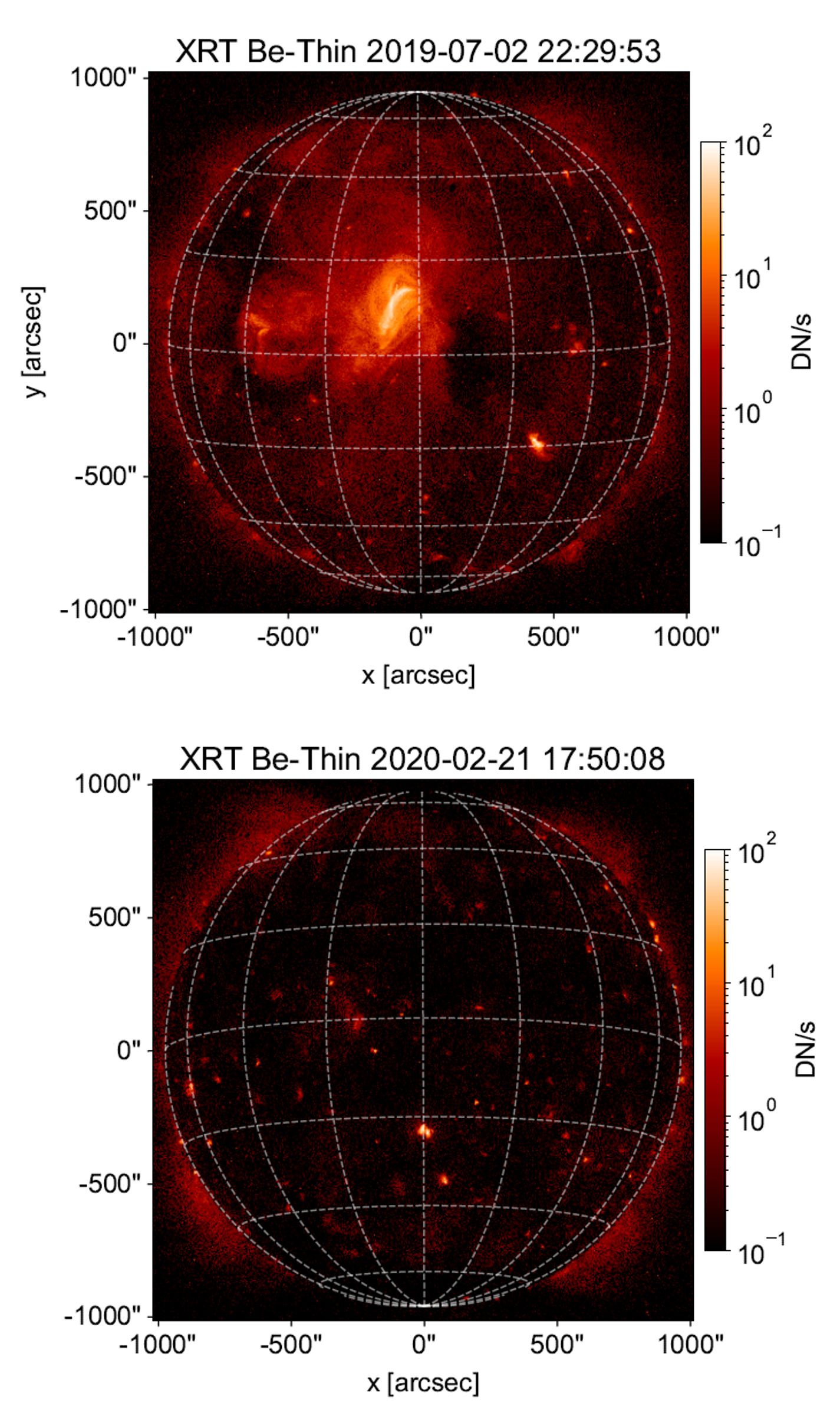}
\caption{Soft X-ray images of the full solar disk from the {\it Hinode} X-Ray Telescope\cite{2008ApJ...672.1237L,2011A&A...526A..78K}. The 2019 image (top) shows an active region near the disk center. The image at the bottom, during the NuSTAR observations, shows only a few X-ray bright points.
} 
\label{fig:corona_pic}
\end{figure}

 From the 3D PSI simulation we define our reference model for the coronal $B_{\perp}$ as follows. 
We average $B_{\perp}$ over the signal region, i.e. the disk within a radius of $0.1\,R_{\odot}$. We repeat this procedure 120 times, corresponding to a complete solar rotation, with the PSI model rotated by an azimuthal angle of $3^\circ$ in each iteration. At each altitude $h$, the median of the resulting 120 values, $\langle |B_\perp|\rangle$, is computed and corresponds to our baseline model used for the calculation of the axion-photon limit. In addition, we compute the lower 5, 10 and 15 percentiles of the distribution. These magnetic field profiles are shown in Fig.~\ref{fig:models}.

To validate this procedure, we compare these results with the prediction of the PFSS model~\cite{pffs6}.
The PFSS model constructs a global magnetic field by assuming a potential field, meaning that it does not incorporate currents threading through the corona as they physically would. Instead, it approximates these effects by setting a boundary condition where the magnetic field becomes purely radial at a distance of $2.5\,\rm{R_\odot}$. This ``source surface" creates artificial surface currents that adjust the internal field to reasonably reflect the influence of photospheric magnetic sources (such as active regions and sunspots) and permits a radial field component, allowing for a representation of the solar wind. 
These publicly accessible models, although not as advanced as the PSI model, allow us to examine the evolution of the coronal magnetic field between the PSI model date and the NuSTAR observation period.
To this end, we compute $|B_{\perp}|$ using the PFSS model specific for the time of the NuSTAR observation. As shown in Fig.~\ref{fig:models} it align closely with the fiducial (median) PSI model described before, strengthening confidence in our results. Instead, the more conservative magnetic field profiles corresponding to the lower 5, 10 and 15 percentiles of the PSI model distribution lie significantly below the PFSS model.

The PSI simulation lacks sufficient resolution to resolve the magnetic field below $h\sim 0.1R_\odot$, while the model for the photosphere from~\cite{Rempel_2014} is reliable only for altitudes $h<400$~km. This leaves a gap in the precise understanding of the magnetic field between these two regions. To bridge this, we applied a simple power-law interpolation, depicted by the dotted line in the upper part of Fig.~\ref{fig:models}. This interpolation estimates a perpendicular magnetic field strength of approximately 1.4~G at an altitude of 10,000~km. 
We notice that this intermediate magnetic field is relevant only for axion masses $\gtrsim 10^{-4}$~eV --
for which the axion-photon conversion process loose coherence in these regions, see Fig.~\ref{fig:prob_of_h}.

\begin{figure}[b!]
 \centering
    \includegraphics[width=0.7\textwidth]{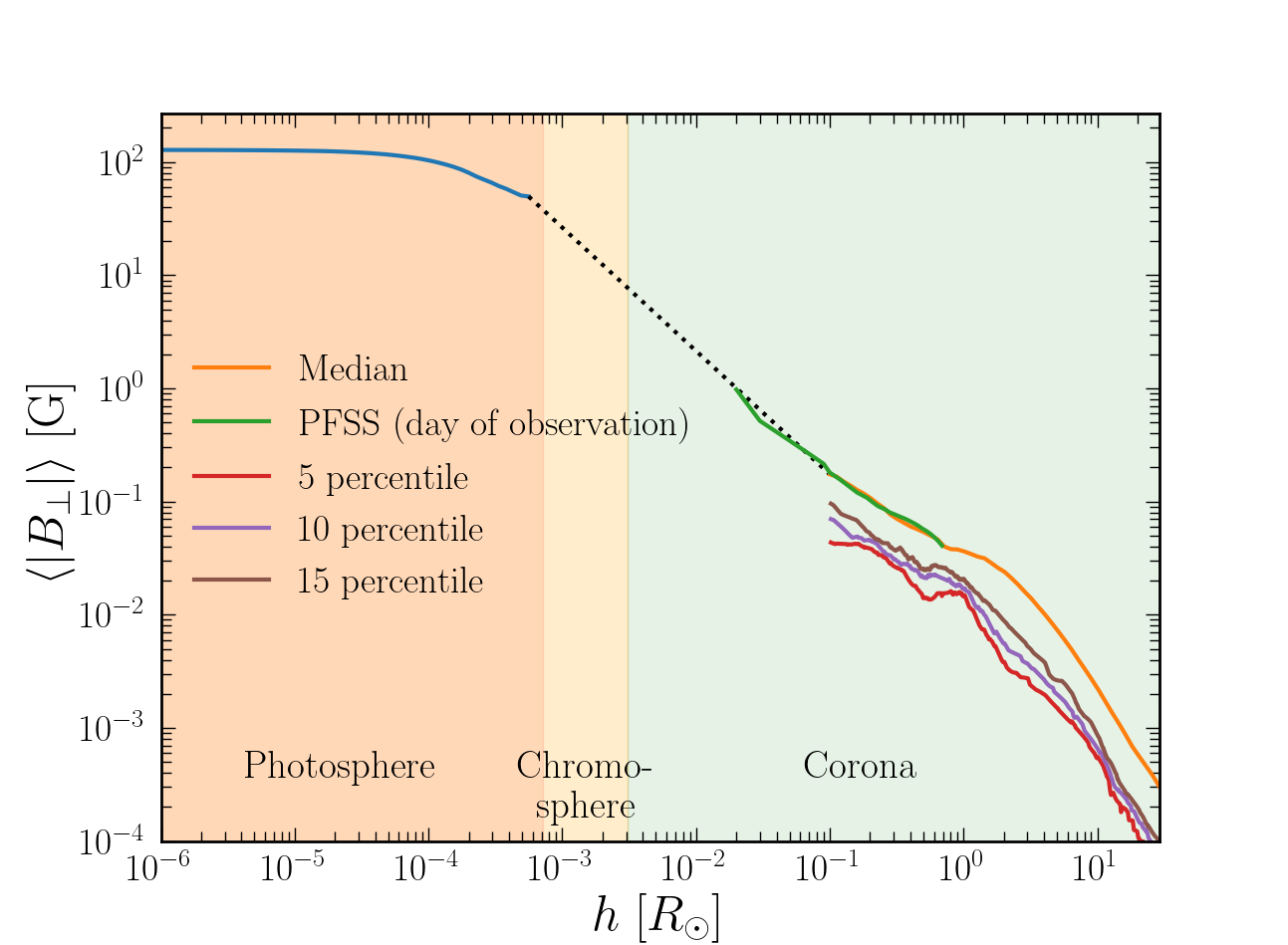}    
    \caption{Our fiducial magnetic field model (blue, dashed black, and orange) derived from the PSI simulation for the 2019 eclipse in comparison to the PFSS model for the day of NuSTAR observations (green). The two models are in excellent agreement in the range of altitudes where the PFSS result is reliable. In red, purple and brown, the 5, 10 and 15 lower percentiles, respectively for the intensity of the solar corona magnetic fields during the period of our observations.}
    \label{fig:models}
\end{figure}

\subsection{Expected X-ray signal} \label{sec:convprob}

To calculate the axion-photon conversion probability, we utilize the components outlined in previous sections and perform a numerical integration of $P_{a\rightarrow\gamma}$ (see Eq. 2 of the main manuscript) up to an altitude of $h=29 R_{\odot}$, the upper boundary of our model. To illustrate the contributions from various layers within the solar atmosphere, Fig.~\ref{fig:prob_of_h} presents the conversion probability as a function of altitude above the photosphere, assuming conversion halts at altitude $h$ and the resulting photons are subsequently detected by NuSTAR. This probability is shown for several axion masses at an energy of $E=4$~keV, with a coupling constant $\gag=10^{-10}$~GeV$^{-1}$.

\begin{figure}[!t]
\centering
    \begin{minipage}{0.6\textwidth}
  \centering
  \includegraphics[width=0.98\linewidth]{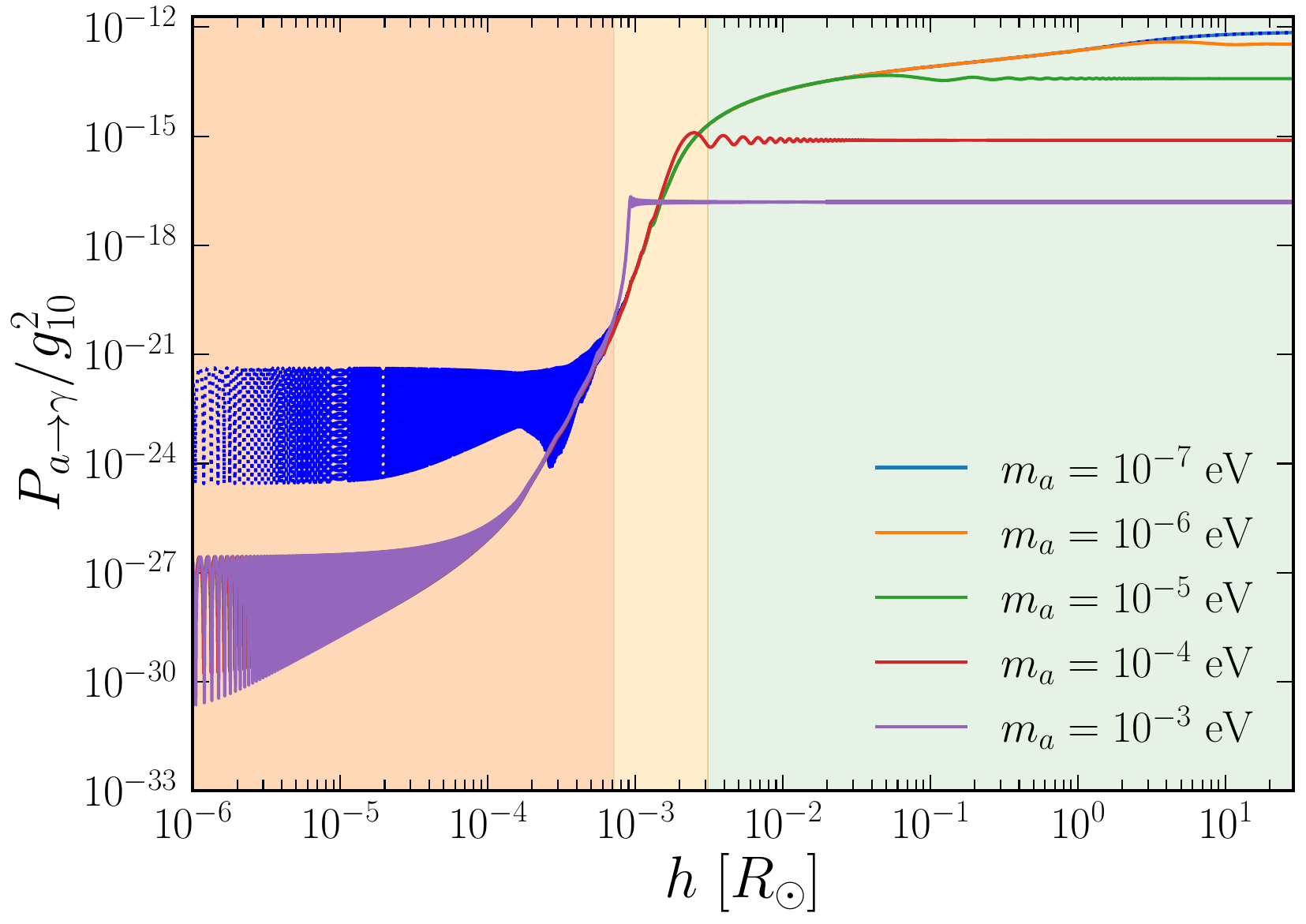}
\end{minipage}
\hspace{0.25cm}
\begin{minipage}{0.6\textwidth}
  \centering
  \includegraphics[width=0.98\linewidth]{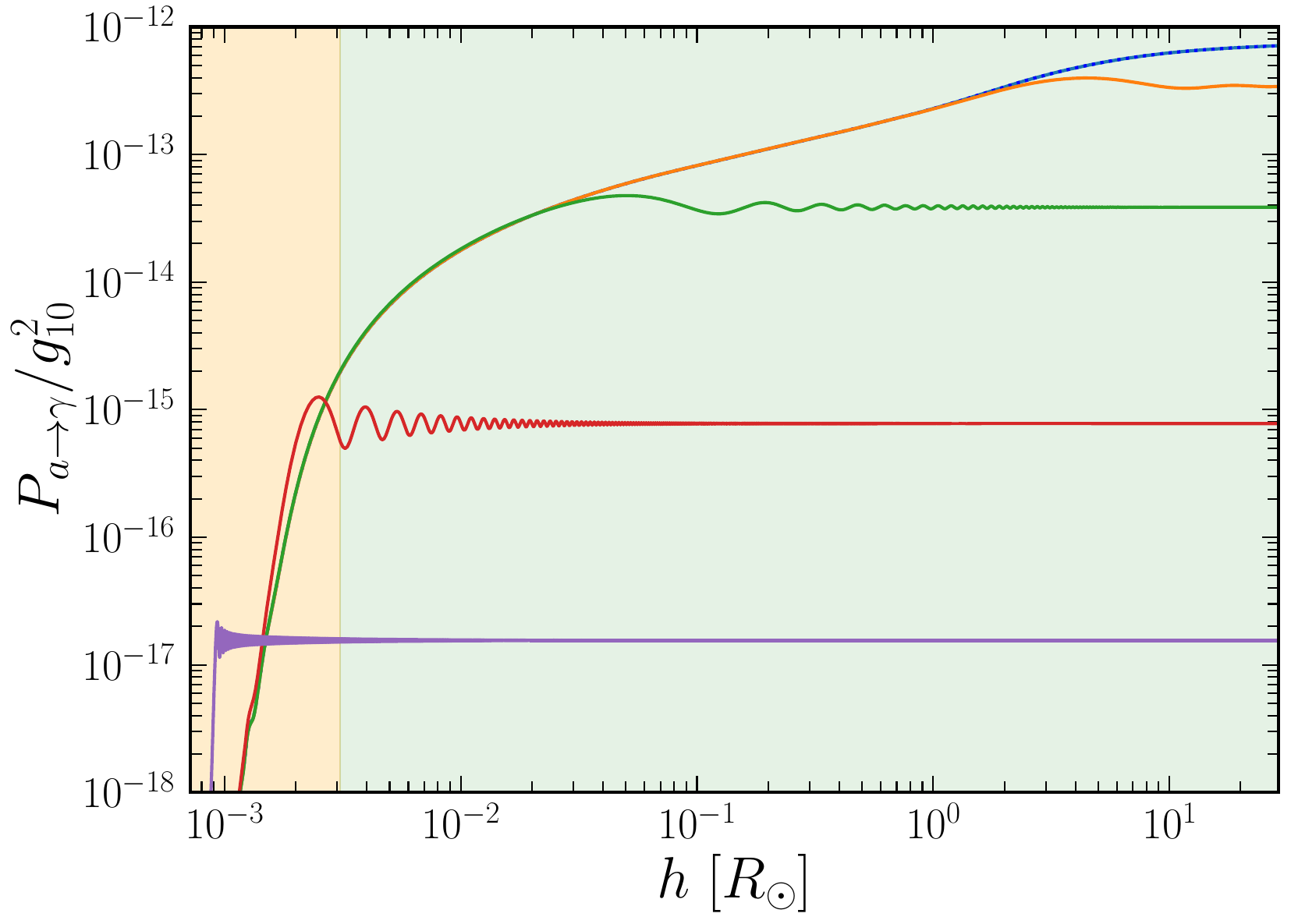}
\end{minipage}
\caption{\textbf{Top:} Overview of the conversion altitude for different axion masses ranging from $10^{-7}$ to $10^{-3}\,\rm{eV}$ and an energy of 4~keV. Curves for lower axion masses are not shown because they would be superimposed to the curve for $m_a=10^{-7}$~eV. This saturation is reflected in our X-ray spectrum  (see Fig.~\ref{fig:mass_dependence}) and bound becoming independent of the axion mass from $m_a \lesssim 10^{-7}$~eV. The dotted blue line shows the conversion probability for $m_a = 10^{-7}\,\rm{eV}$ where absorption effects have been turned off. Absorption is only significant at low altitudes, where the conversion probability is negligibly small. At altitudes larger than $10^{-3}~R_\odot$ the dotted line is superimposed with the solid blue line. The photosphere, chromosphere, and corona regions of our Sun's atmosphere are shown in orange, yellow, and green, respectively.  \textbf{Bottom:} Closeup of the chromosphere and corona, where the probability of conversion is maximized. }\label{fig:prob_of_h}
\end{figure}

At altitudes $h < 10^{-4}R_\odot$,  the conversion probability remains low due to a lack of coherence ($qh\gg 1$), resulting in suppressed X-ray emission. The probability begins to increase as the plasma density decreases, enabling more effective conversion. When the plasma density drops sufficiently and the axion mass is negligible compared to the plasma frequency, the phase $\varphi = \int dh\, q$ becomes approximately constant, 
effectively factoring out. Simultaneously, reduced absorption enhances the conversion process. When the plasma frequency reaches the axion mass, i.e., at resonance, $q= (\omega_{p}^2 - m_a^2)/2E$ changes sign and becomes dominated by the axion mass. Beyond this point, the complex phase oscillation $\varphi$ becomes significant, particularly for higher axion masses that lead to shorter oscillation wavelengths. This effect, coupled with the decreasing magnetic field, flattens the growth of conversion probability. Since the plasma frequency remains above $10^{-9}$~eV within the altitude range of our model (see Fig. 2 of the main text), resonance is not met for $m_a\lesssim 10^{-9}\,{\rm eV}$, and conversion probability flattens as the coronal magnetic field diminishes. For lower axion masses, the photon production rate, $dP/dh$, peaks in the mid-to-upper corona. Photon production at even greater heights up to 1 AU is conservatively neglected.

\begin{figure}[!t]
    \centering
    \includegraphics[width=0.7\textwidth]{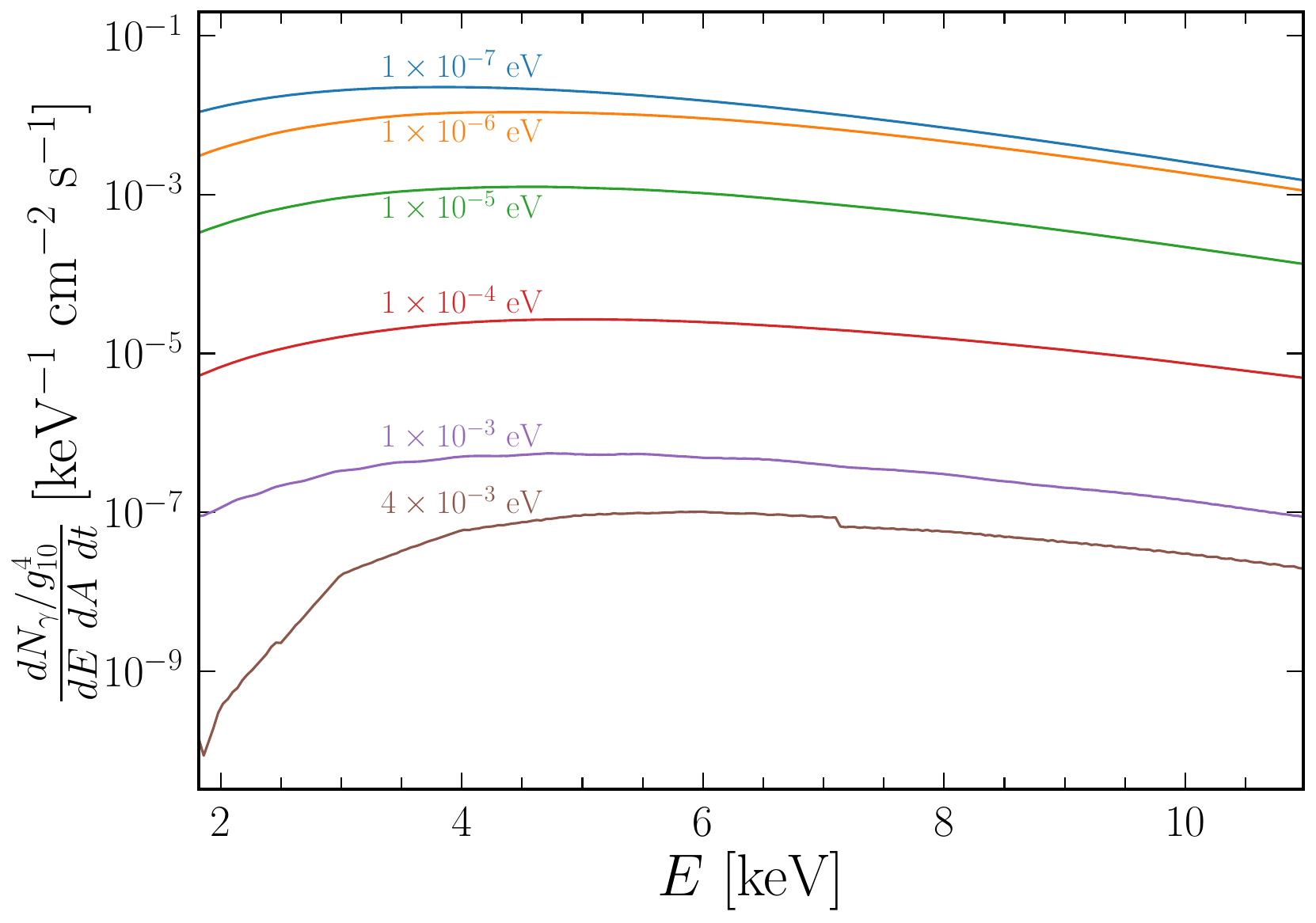}   
    \caption{X-ray spectra reaching NuSTAR satellite depending on axion mass. Notice how the K-shell absorption due to iron presence in our Sun happens for considerably large axion masses $m_{a}\gtrsim 10^{-3}\,\rm{eV}$. These axions fulfill coherence of conversion in deeper layers than the chromosphere, and their overall flux contribution is considerably smaller (4 orders of magnitude less) than in the case of $m_{a}\sim 10^{-7}\,\rm{eV}$.}
    \label{fig:mass_dependence}
\end{figure}
Fig.~\ref{fig:mass_dependence} presents the expected photon flux from our signal region for different axion masses. For the largest axion mass shown, $m_a = 4\times 10^{-3}$~eV, the effect of photoelectric absorption is clearly apparent. Specifically, increased absorption is observed for X-ray photons with energies above the K-edge of iron at approximately $\sim 7.1$~keV. However, for the axion masses where the coherence is completely fulfilled, $m_{a}\lesssim 5\times 10^{-4}$~eV , the presence of K-edges becomes negligible.
\begin{figure}[!b]
    \centering
    \includegraphics[width=0.7\textwidth]{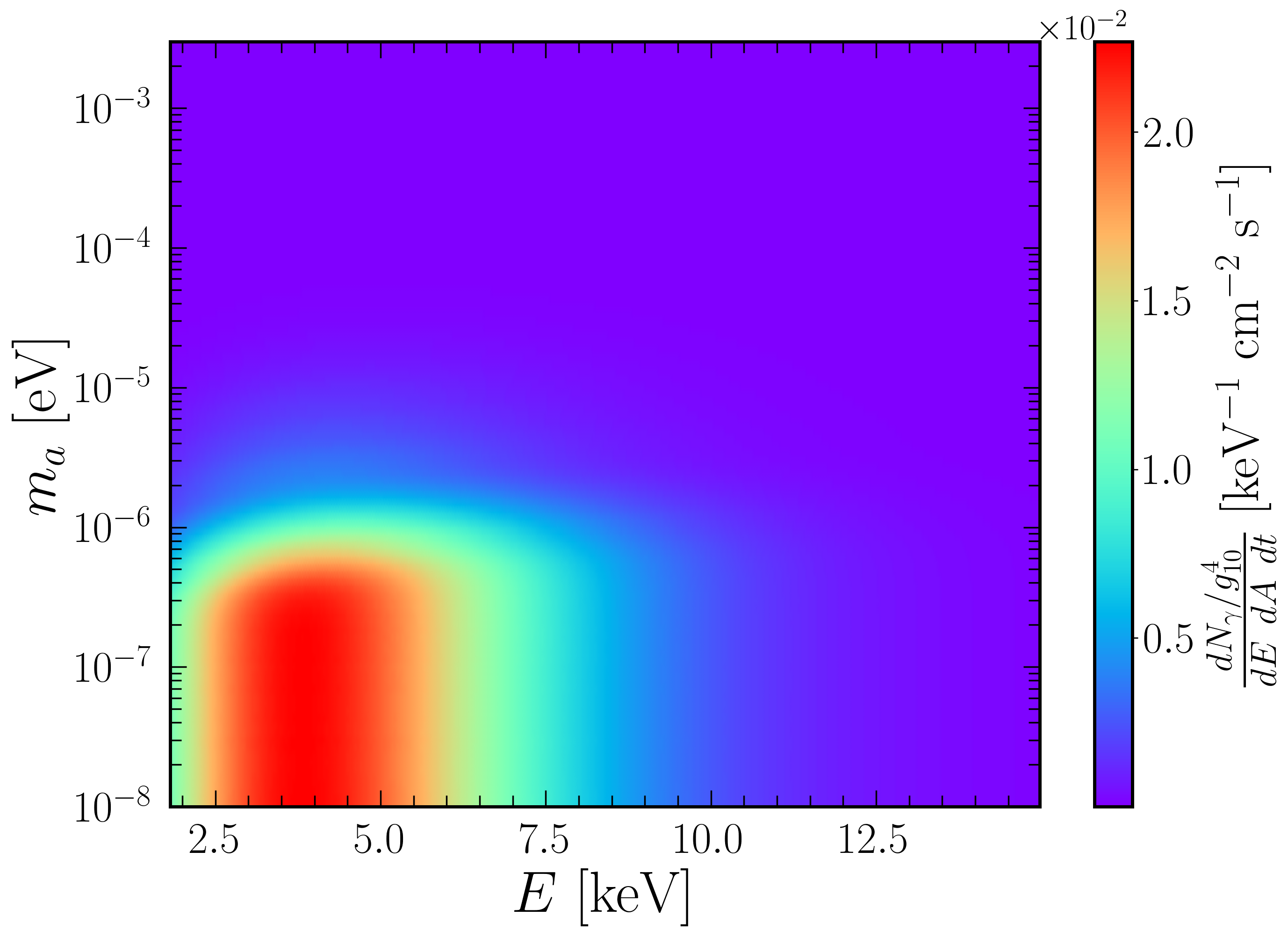}
    \caption{Number of photons due to axion conversion per unit energy, area and time, for an axion-to-photon coupling $g_{10}=1$, where $g_{10}=\gag/(10^{-10}~\mathrm{GeV}^{-1})$.} 
    \label{fig:axion_spectrum}
\end{figure}
Fig.~\ref{fig:axion_spectrum} shows the expected photon flux from our signal region $R<0.1 R_\odot$. We can see that, as expected, the signal peaks at around 4 keV. Fig.~\ref{fig:prob_of_h} shows that for masses larger than $10^{-6}$~eV, the conversion probability stops growing at lower and lower altitude in the solar atmosphere, leading to a reduced photon flux. This explains the drop in the photon production for $m_a \gtrsim 10^{-6}$~eV seen in Figure~\ref{fig:axion_spectrum}. 

%
%
\section{NuSTAR's Solar Observation Campaigns}\label{sec:solar_NuSTAR}
Since 2014, NuSTAR has conducted several observations of the Sun, with an overview of these observations and quick-look plots available in~\cite{iain_github}. These observations have primarily focused on weakly flaring active regions, often located away from the disk center and therefore, not ideal for axion searches. Even when NuSTAR targets quieter regions of the solar disk, stray X-rays—referred to as ``ghost rays"—can still be detected hindering low-event studies. These rays result from a single reflection in the two-bounce optics and can originate from active regions outside the $10'\times10'$ field of view~\cite{Grefenstette_2016}. As a result, for axion searches, the entire solar disk must be relatively quiet.
The minimum of the 11-year solar activity cycle provided a unique opportunity to study the emission from the quiet Sun, and NuSTAR conducted several campaigns targeting this period from 2018 to 2020. Among these, the 21-Feb-2020 campaign offers the best data for axion studies: This campaign involved a long dwell at the disk center during a period of very low solar activity, with 9 orbits of observations yielding approximately 23.8~ks  and 24.9~ks of data from NuSTAR's telescope modules $A$ and $B$, respectively. Despite the quiet conditions, faint X-ray features, known as X-ray bright points, are present in the solar atmosphere. A detailed analysis of these features is presented in~\cite{sarah24}, where their spectra are found to be consistent with optically-thin thermal emission sources. The effective temperature of these emissions is typically $\leq$3~MK, with occasional brief brightening reaching closer to 4 MK. Such sources produce X-ray spectra that steeply decrease with photon energy, making them nearly undetectable above 4
~keV amidst the background noise. While other NuSTAR observations from this solar minimum exist, they were part of mosaics covering the entire solar disk, contributing only about 100 extra seconds of data. 

With the next solar minimum expected in 2030, the current dataset represents the best X-ray satellite observation for solar axion analysis in the near future.
\subsection{NuSTAR data processing for solar observations}\label{sec:processing}
NuSTAR collected quiet Sun data over nine spacecraft orbits between 04:51 GMT and 22:41 GMT on February 21, with each orbit providing about one hour of observation. The $10'$ square field of view (FoV) covered $\sim18\%$ of the solar disk. While the satellite's orbit was selected to minimize the impact of passage through the South Atlantic Anomaly (SAA)~\cite{2013ApJ...770..103H}, most observations were affected by observational gaps or dead times due to such passages. The SAA is a geomagnetic anomaly where the weakened magnetic field over the southern Atlantic Ocean exposes spacecraft in low-Earth orbit, like $\rm{NuSTAR}$, to Earth's inner radiation belt, extending down to the upper atmosphere~\cite{Dessler1959}. As the Sun drifted slowly across the FOV during the observations, we adjusted the coordinates of the solar center in 10 to 15-minute intervals (segments) to minimize the effects of this drift. Separate data files were extracted for each orbit segment and the details of each orbit can be found in Table~\ref{tab:data_sets}.
 
\begin{table}[!t]
\renewcommand{\arraystretch}{2.0}
    \centering
    \begin{tabular}{c c c c c} 
    \toprule
        \textbf{No.} & \textbf{ID} & \textbf{On-Target} & \textbf{Time} & \textbf{Correction}\\ 
          & & (GMT) & (min) & \\\toprule
         1 & 80512218001& 05:16:13 - 06:15:48 & 59.6 & 0.934\\ \hline
         2 & 80512220001& 08:38:10 - 09:29:06 & 50.9 & 0.962\\ \hline 
         3 & 80512221001& \shortstack{ \\ 10:06:09 - 10:09:56 \\10:22:18 - 11:05:44} & 47.2 & 0.965\\ \hline
         4 & 80512222001& \shortstack{ \\ 11:42:48 - 11:51:42 \\12:06:02 - 12:42:23} & 45.3 & 0.981\\ \hline
         5 & 80512223001& \shortstack{ \\ 13:19:26 - 13:33:56 \\13:49:18 - 14:19:02 } & 44.2 & 0.974\\ \hline
         6 & 80512224001& \shortstack{ \\ 14:56:05 - 15:16:44 \\15:32:30 - 15:55:40 } & 43.9 & 0.960\\ \hline
         7 & 80512225001& \shortstack{ \\ 16:32:44 - 17:00:28 \\17:15:26 - 17:32:19 } & 44.6 & 0.981\\ \hline
         8 & 80512226001& \shortstack{ \\ 18:09:22 - 18:46:09 \\18:57:58 - 19:08:58 } & 47.8 & 0.999\\ \hline
         9 & 80512228001& 21:22:39 - 22:22:15 & 59.6 & 0.945\\ \bottomrule
    \end{tabular}
    \vspace{0.5cm}
    \caption{Basic information for each orbit of solar data collected on February 21, 2020, with NuSTAR. Orbits with two time intervals were interrupted by a period of passage through the SAA. The last column shows the stray light scaling factor by which the background collected in the outer annulus region was multiplied to account for the known gradient of stray light background across the chip.} 
    \label{tab:data_sets}
\end{table}
The data were processed using NuSTAR Data Analysis Software (\texttt{NuSTARDAS}) version 2.1.4 to generate scientific products. \texttt{NuSTARDAS} is integrated within the \texttt{NASA-HEASARC HEASoft} software framework (version 6.34), which was downloaded along with the \texttt{CALDB} calibration database (version 20240325)~\cite{Perri2020}. \texttt{XSPEC} tool \cite{arnaud1999xspec} version 12.13.1 was used to read the spectra and export the data into ASCII format. The data processing involved three main steps: calibration, screening, and product extraction.
All 32 grades of NuSTAR data were collected without filtering, with each grade corresponding to a distinct pattern of energy deposition across the detector pixels~\cite{Perri2020}. This resulted in level 0 data, which consists of raw telemetry packets that were then formatted into Level 1 FITS files prior to calibration. We processed the Level 1 data using the software module {\it nupipeline} to generate Level 1a calibrated files and Level 2 calibrated, cleaned files.

The calibration process includes several steps, such as {\it nuhotpix} (for identifying hot pixels), {\it nucalcpha} (for assigning energy and grade to each event), and {\it nucoord} (for converting raw coordinates into detector and sky coordinates). Screening, also referred to as cleaning, involves additional processes like {\it nucalcsaa} (for calculating SAA passages), {\it nulivetime} (for correcting dead times in the event files), and {\it nusplitsc} (for splitting the cleaned event files into separate files based on the spacecraft's switching between cameras used for attitude determination).
 
The initial processing was followed by selection of the source and background regions. The source region is a circle centered at the solar disk center with radius $0.1 R_{\odot}$ (1.6'), and the background region is an annulus from $0.15$ to $0.30$ solar radius.  Wedges were occasionally excluded from the background annulus where examination of the energy range from 3.5-4.0~keV showed an X-ray bright point.  Generally, no obvious excess was visible above 4.0 keV, but we nevertheless removed these regions in case their aggregate contained a slight excess not clearly visible in any one image.  This method avoids bias by not using the data that actually contribute to the analysis ($>$4.0~keV) to decide what areas to excise. The editing of the annulus region was done separately for each 10--15~min sub-interval orbit. Fig.~\ref{fig:regions} shows the signal and background regions superimposed on data for all the time intervals of one orbit. 

\begin{figure}[t!]
\centering
\includegraphics[width=0.5\textwidth]{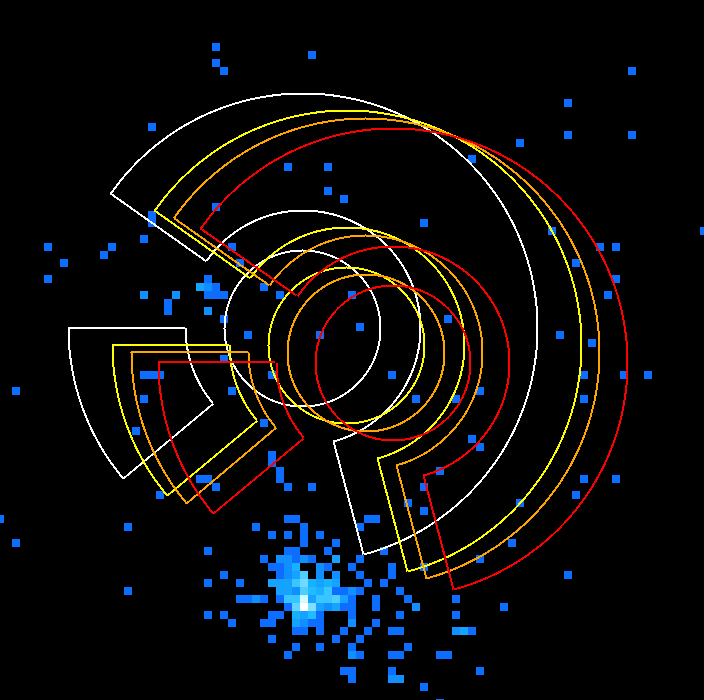}  
\caption{Data extraction regions for orbit 80512226001, telescope module A.  Colour scale: counts from 3.5-4.0 keV, binned to 4x4 detector pixels; darkest blue = 1 count, white = 53 counts.  The inner circles are the extraction regions for the disk center source (1.6’ radius), and the outer annulus segments are the background extraction region, avoiding the solar X-ray bright point at the bottom of the image and the faint one at the upper left. The bright spot at the bottom coincides with the brightest in the bottom part of Fig.~\ref{fig:corona_pic}.   The contours running from red to white were used sequentially during this orbit to track the motion of the Sun on the sky.
} 
\label{fig:regions}
\end{figure} 

The first part of the entire {\it nuproducts} product extraction sequence includes the following functions: {\it nuproducts} extracting refined PHA files and light curves by slicing out our chosen time intervals and the shapes of our supplied region files; `numkarf', creating ancillary response files (ARFs) which detail the telescope modules' responses as a function of position and energy; and `numkrmf', creating redistribution matrix files (RMFs) which detail the detectors' response as a function of photon energy. We executed {\it nuproducts} for each source and background region by supplying directories, IDs, event files, region files, time interval files, and setting automatic background scaling and extraction to `no'.
The second part of the {\it nuproducts} sequence is called `nubackscale', where we extracted our own background scaling variables separately. These variables define what proportion our source and background regions are of the entire FOV without hot/bad pixels and detector gaps. We executed `nubackscale' by supplying the PHA files of both the source and background from the previous step as well as other reference files pertaining to the observation ID. This added a \ttfamily{BACKSCAL}\rmfamily~keyword to new PHA files, which we used in follow-on steps.

``Stray light" (cosmic diffuse X-rays that enter the detector without encountering the optics) is our dominant source of background and has a distinctive gradient pattern across the detector plane, based on variable shadowing by the spacecraft structure and optics~\cite{Wik_2014}.  Using the {\it nuskybgd} package, we determined the effect of this pattern on each of our source and background annulii regions. The \texttt{XSPEC} software tool (version 12.13.1) was used as well to obtain spectra.

\section{Sensitivity of the results to model uncertainties. Systematic studies}
\label{sec:results}\label{app:systematics}
The sources of systematic uncertainty in our bounds include the solar magnetic field, the solar axion flux, and the NuSTAR background. To assess how these factors influence our results, we first explore how the bounds vary with changes in the profile shape and normalization of the coronal magnetic field model. Our photospheric model is already a conservative estimate, as it represents the inter-network regions of the quiet Sun, excluding contributions from network and active regions. To estimate the uncertainty arising from the shape of the coronal magnetic field profile, we recompute our bound in the limit of vanishing axion mass using 120 profiles, each obtained by rotating the model by $3^\circ$ increments (see section on the solar magnetic field)
and to isolate the dependence on the profile shape, we rescale each profile so that the average magnetic field strength $\langle|B_\perp|\rangle$ at an altitude of $h=0.1R_\odot$ matches that of our fiducial model. This also aligns with the PFSS model for the day of observation. We then recompute the bound for each of the 120 profiles and report the systematic uncertainty in Table~\ref{tab:systematic}, expressed as the percentage variation in the bound. 

The uncertainty arising from the normalization of the coronal magnetic field is estimated from the variance of the PFSS model over the perpendicular disk of radius $0.1R_\odot$ at an altitude of $h=0.1R_\odot$. In our reference model, the value of $\langle|B_\perp|\rangle$ at this altitude is 0.174~G. From the PFSS model, we estimate the $1\sigma$ range to be $0.155~\mathrm{G}<|B_\perp|<0.190~\mathrm{G}$. The variance at this altitude is larger than that at higher altitudes ($h>0.1R_\odot$), and $h=0.1R_\odot$ is the lowest altitude at which we use the coronal model from PSI. To estimate the corresponding uncertainty, we rescale our fiducial magnetic field by an overall factor so that its value at $h=0.1R_\odot$ is either 0.155~G or 0.190~G. The resulting change in the bound is reported in Table~\ref{tab:systematic} as a percentage variation.
\begin{table}[t!]
    \centering
    \begin{tabular}{l c c}
         \toprule
         \textbf{Quantity} &  \multicolumn{2}{c}{\textbf{Systematic effect on $g_{a\gamma}$}}\\ \hline
         Coronal $\langle|B_\perp|\rangle$ shape  & $-18\%$ &$+26\%$ \\ \hline
         Coronal $\langle|B_\perp|\rangle$ normalization  & $-11\%$ & $+9.0\%$\\ \hline
         Solar axion flux& $-1.5\%$ &$+1.5\%$  \\ \hline
         NuSTAR background & $-1.5\%$ & $+1.6\%$ \\  \bottomrule
    \end{tabular}
    \caption{Sources of systematic uncertainty and their effect on our bound for vanishing axion mass. }
    \label{tab:systematic}
\end{table}
Another source of uncertainty is the solar axion flux, that has been recently revisited in~\cite{Hoof:2021mld}, estimating a flux uncertainty of 6\% due to the solar model. This rescaling of the axion flux has a minor effect on our bounds, as shown in Table~\ref{tab:systematic}. Additionally, we considered the potential impact of the Cosmic X-ray Background (CXB) as a possible source of background variability. The dominant background component at the energies where we expect an axion signal is the aperture Cosmic X-Ray Background (aCXB), which could present spatial variations across the NuSTAR detectors, potentially leading to an imperfect subtraction of this component in our analysis. Based on previous systematic studies of the NuSTAR instrument~\cite{Rossland_2023}, we account for background uncertainties of $\pm 1.5\%$. The final line of Table~\ref{tab:systematic} shows how our bound for $m_a\to 0$ is affected by this uncertainty in the photon counts within the annular region if the background estimation is adjusted by this factor.

\section{Comparison to astrophysical bounds}
\label{sec:comp}

In addition to the experimental limits presented in the main body, there are several other competitive, yet more model-dependent, astrophysical bounds on the axion-to-photon coupling $g_{a\gamma}$ that have been reported in the literature. These limits are shown in Fig.~\ref{fig:exclusion_SM}.

Studies in Refs.~\cite{Li_2022, Reynolds_2020, Abramowski_2013, Ajello_2016, Davies_2023, Jacobsen:2022swa,Li:2024zst} focus on detecting the imprint of axion-photon mixing in the energy spectra of extra-galactic gamma-ray and X-ray sources. 
Other notable targets for constraining axion-photon conversion include super-giant stars~\cite{Xiao:2020pra}, stars in external galaxies~\cite{Ning:2024ek}, magnetic white dwarfs~\cite{Dessert_2022_1, Dessert_2022_2}, pulsars~\cite{Noordhuis:2022ljw}, super star clusters~\cite{Dessert_2020}, and the supernova SN1987A~\cite{Hoof_2023,Manzari:2024jns}. However, although these systems provide excellent opportunities for axion searches, they come with the hurdle of estimating the size of uncertainties for the intergalactic magnetic field and plasma density models, which significantly affect the derived bounds.

Other astrophysical bounds shown 
in Fig.~\ref{fig:exclusion_SM}, are derived from neutron stars~\cite{Foster:2022fxn, Battye:2023oac} or the explosion of hypothetical axion stars~\cite{Escudero:2023vgv} and, are based on the assumption that all dark matter consists of axions. In contrast, the limits obtained in this work, as well as those discussed earlier, do not rely on such an assumption. 
\begin{figure}[t!]
    \centering
    \includegraphics[width=0.7\textwidth]{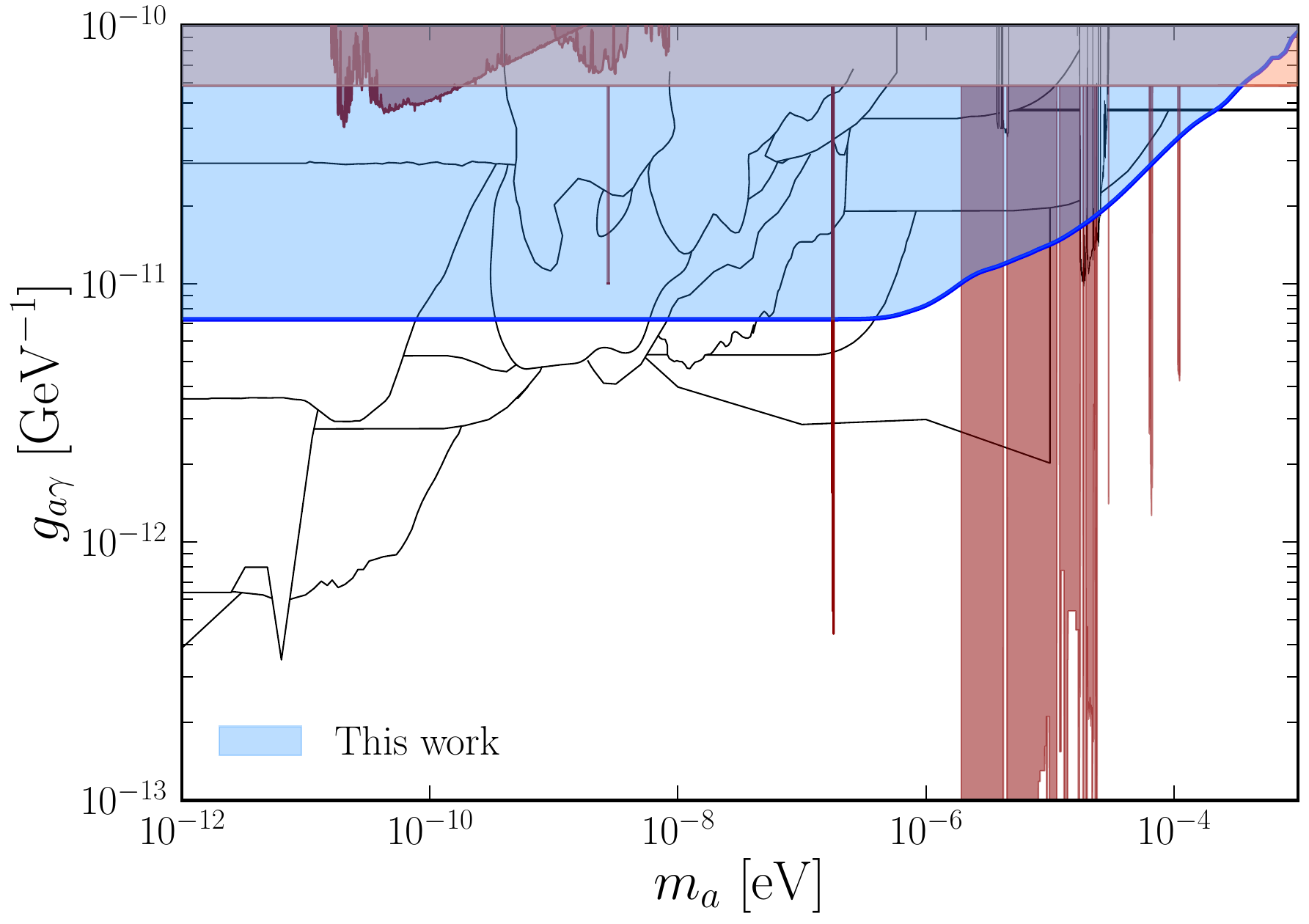}
    \caption{NuSTAR's 95\% CL upper limit on the axion–photon coupling strength $g_{a\gamma }$ from our work (blue line).
    Bounds from haloscopes~\cite{Asztalos_2010, Du_2018, Braine_2020, Bartram_2021, Boutan_2018, Bartram_2023, PhysRevD.42.1297, Zhong_2018, Backes_2021, haystaccollaboration2023new, Lee_2020, Jeong_2020, Kwon_2021, Lee_2022, Kim_2023, Yi_2023, Yang_2023, kim2023experimental, PhysRevX.14.031023, Quiskamp_2022, Abeln:2021, McAllister_2017, Alesini_2019, Alesini_2021, Di_Vora_2023, Devlin:2021fpq,Ouellet_2019, Salemi_2021, Crisosto:2019fcj} are shown in dark red while the CAST helioscope constraint~\cite{Andriamonje_2007, cast_jcap2009,CAST_PRL2011, PhysRevLett.112.091302, CAST_nature, CAST_PRL04} is in light red.
    The dark green area corresponds to the limits from globular clusters derived from the R parameter~\cite{Ayala_2014}, while the thin black lines are the astrophysical bounds derived in~\cite{ Li_2022, Reynolds_2020, Abramowski_2013, Ajello_2016, Davies_2023, Jacobsen:2022swa,Li:2024zst,Xiao:2020pra,Ning:2024ek,Dessert_2022_1, Dessert_2022_2,Noordhuis:2022ljw,Dessert_2020,Hoof_2023,Manzari:2024jns}.
    The astrophysical bounds in purple from~\cite{Foster:2022fxn, Battye:2023oac,Escudero:2023vgv} rely on the assumption that all the dark matter is in the form of axions. The code to generate this plot was adapted from \url{https://github.com/cajohare/AxionLimit}.}
    \label{fig:exclusion_SM}
\end{figure}
Lastly, 
we also present in Fig.~\ref{fig:exclusion_SM} limits from globular clusters, derived from the R parameter~\cite{Ayala_2014}, which measures the ratio of horizontal branch stars to red giant stars. These bounds are comparable to those obtained from the ${\rm R}_2$ parameter~\cite{Dolan_2022}, which measures the ratio of stellar populations on the asymptotic giant branch to those on the horizontal branch.

As illustrated in Fig.~\ref{fig:exclusion_SM}, many of the methods discussed in this section show promising potential for axion detection. However, the associated limits often come with larger uncertainties than those derived in this study. 
As mentioned earlier, these uncertainties arise primarily from the complexity of modeling axion production and conversion in the far away astrophysical objects and intergalactic environments studied.  
In this context, the search presented in this paper offers a distinct advantage: The Sun provides an environment that is studied in great detail, allowing for a robust assessment of the model uncertainties, and leading to more reliable bounds on axions.

%% file: main_PRL_v3.0_arXiv.bbl
\begin{thebibliography}{114}%
\makeatletter
\providecommand \@ifxundefined [1]{%
 \@ifx{#1\undefined}
}%
\providecommand \@ifnum [1]{%
 \ifnum #1\expandafter \@firstoftwo
 \else \expandafter \@secondoftwo
 \fi
}%
\providecommand \@ifx [1]{%
 \ifx #1\expandafter \@firstoftwo
 \else \expandafter \@secondoftwo
 \fi
}%
\providecommand \natexlab [1]{#1}%
\providecommand \enquote  [1]{``#1''}%
\providecommand \bibnamefont  [1]{#1}%
\providecommand \bibfnamefont [1]{#1}%
\providecommand \citenamefont [1]{#1}%
\providecommand \href@noop [0]{\@secondoftwo}%
\providecommand \href [0]{\begingroup \@sanitize@url \@href}%
\providecommand \@href[1]{\@@startlink{#1}\@@href}%
\providecommand \@@href[1]{\endgroup#1\@@endlink}%
\providecommand \@sanitize@url [0]{\catcode `\\12\catcode `\$12\catcode
  `\&12\catcode `\#12\catcode `\^12\catcode `\_12\catcode `\%12\relax}%
\providecommand \@@startlink[1]{}%
\providecommand \@@endlink[0]{}%
\providecommand \url  [0]{\begingroup\@sanitize@url \@url }%
\providecommand \@url [1]{\endgroup\@href {#1}{\urlprefix }}%
\providecommand \urlprefix  [0]{URL }%
\providecommand \Eprint [0]{\href }%
\providecommand \doibase [0]{https://doi.org/}%
\providecommand \selectlanguage [0]{\@gobble}%
\providecommand \bibinfo  [0]{\@secondoftwo}%
\providecommand \bibfield  [0]{\@secondoftwo}%
\providecommand \translation [1]{[#1]}%
\providecommand \BibitemOpen [0]{}%
\providecommand \bibitemStop [0]{}%
\providecommand \bibitemNoStop [0]{.\EOS\space}%
\providecommand \EOS [0]{\spacefactor3000\relax}%
\providecommand \BibitemShut  [1]{\csname bibitem#1\endcsname}%
\let\auto@bib@innerbib\@empty
\bibitem [{\citenamefont {{Peccei}}\ and\ \citenamefont
  {{Quinn}}(1977)}]{1977PhRvL..38.1440P}%
  \BibitemOpen
  \bibfield  {author} {\bibinfo {author} {\bibfnamefont {R.~D.}\ \bibnamefont
  {{Peccei}}}\ and\ \bibinfo {author} {\bibfnamefont {H.~R.}\ \bibnamefont
  {{Quinn}}},\ }\href {https://doi.org/10.1103/PhysRevLett.38.1440} {\bibfield
  {journal} {\bibinfo  {journal} {Phys. Rev. Lett.}\ }\textbf {\bibinfo
  {volume} {38}},\ \bibinfo {pages} {1440} (\bibinfo {year}
  {1977})}\BibitemShut {NoStop}%
\bibitem [{\citenamefont {Weinberg}(1978)}]{PhysRevLett.40.223}%
  \BibitemOpen
  \bibfield  {author} {\bibinfo {author} {\bibfnamefont {S.}~\bibnamefont
  {Weinberg}},\ }\href {https://doi.org/10.1103/PhysRevLett.40.223} {\bibfield
  {journal} {\bibinfo  {journal} {Phys. Rev. Lett.}\ }\textbf {\bibinfo
  {volume} {40}},\ \bibinfo {pages} {223} (\bibinfo {year} {1978})}\BibitemShut
  {NoStop}%
\bibitem [{\citenamefont {Wilczek}(1978)}]{PhysRevLett.40.279}%
  \BibitemOpen
  \bibfield  {author} {\bibinfo {author} {\bibfnamefont {F.}~\bibnamefont
  {Wilczek}},\ }\href {https://doi.org/10.1103/PhysRevLett.40.279} {\bibfield
  {journal} {\bibinfo  {journal} {Phys. Rev. Lett.}\ }\textbf {\bibinfo
  {volume} {40}},\ \bibinfo {pages} {279} (\bibinfo {year} {1978})}\BibitemShut
  {NoStop}%
\bibitem [{\citenamefont {Abbott}\ and\ \citenamefont
  {Sikivie}(1983)}]{ABBOTT1983133}%
  \BibitemOpen
  \bibfield  {author} {\bibinfo {author} {\bibfnamefont {L.}~\bibnamefont
  {Abbott}}\ and\ \bibinfo {author} {\bibfnamefont {P.}~\bibnamefont
  {Sikivie}},\ }\href
  {https://doi.org/https://doi.org/10.1016/0370-2693(83)90638-X} {\bibfield
  {journal} {\bibinfo  {journal} {Physics Letters B}\ }\textbf {\bibinfo
  {volume} {120}},\ \bibinfo {pages} {133} (\bibinfo {year}
  {1983})}\BibitemShut {NoStop}%
\bibitem [{\citenamefont {Preskill}\ \emph {et~al.}(1983)\citenamefont
  {Preskill}, \citenamefont {Wise},\ and\ \citenamefont
  {Wilczek}}]{Preskill:1982cy}%
  \BibitemOpen
  \bibfield  {author} {\bibinfo {author} {\bibfnamefont {J.}~\bibnamefont
  {Preskill}}, \bibinfo {author} {\bibfnamefont {M.~B.}\ \bibnamefont {Wise}},\
  and\ \bibinfo {author} {\bibfnamefont {F.}~\bibnamefont {Wilczek}},\ }\href
  {https://doi.org/10.1016/0370-2693(83)90637-8} {\bibfield  {journal}
  {\bibinfo  {journal} {Phys. Lett.}\ }\textbf {\bibinfo {volume} {B120}},\
  \bibinfo {pages} {127} (\bibinfo {year} {1983})}\BibitemShut {NoStop}%
\bibitem [{\citenamefont {Dine}\ \emph {et~al.}(1981)\citenamefont {Dine},
  \citenamefont {Fischler},\ and\ \citenamefont {Srednicki}}]{Dine:1981rt}%
  \BibitemOpen
  \bibfield  {author} {\bibinfo {author} {\bibfnamefont {M.}~\bibnamefont
  {Dine}}, \bibinfo {author} {\bibfnamefont {W.}~\bibnamefont {Fischler}},\
  and\ \bibinfo {author} {\bibfnamefont {M.}~\bibnamefont {Srednicki}},\ }\href
  {https://doi.org/10.1016/0370-2693(81)90590-6} {\bibfield  {journal}
  {\bibinfo  {journal} {Phys. Lett.}\ }\textbf {\bibinfo {volume} {104B}},\
  \bibinfo {pages} {199} (\bibinfo {year} {1981})}\BibitemShut {NoStop}%
\bibitem [{\citenamefont {Dine}\ and\ \citenamefont
  {Fischler}(1983)}]{Dine:1982ah}%
  \BibitemOpen
  \bibfield  {author} {\bibinfo {author} {\bibfnamefont {M.}~\bibnamefont
  {Dine}}\ and\ \bibinfo {author} {\bibfnamefont {W.}~\bibnamefont
  {Fischler}},\ }\href {https://doi.org/10.1016/0370-2693(83)90639-1}
  {\bibfield  {journal} {\bibinfo  {journal} {Phys. Lett.}\ }\textbf {\bibinfo
  {volume} {B120}},\ \bibinfo {pages} {137} (\bibinfo {year}
  {1983})}\BibitemShut {NoStop}%
\bibitem [{\citenamefont {Jaeckel}\ and\ \citenamefont
  {Ringwald}(2010)}]{Jaeckel:2010ni}%
  \BibitemOpen
  \bibfield  {author} {\bibinfo {author} {\bibfnamefont {J.}~\bibnamefont
  {Jaeckel}}\ and\ \bibinfo {author} {\bibfnamefont {A.}~\bibnamefont
  {Ringwald}},\ }\href {https://doi.org/10.1146/annurev.nucl.012809.104433}
  {\bibfield  {journal} {\bibinfo  {journal} {Ann. Rev. Nucl. Part. Sci.}\
  }\textbf {\bibinfo {volume} {60}},\ \bibinfo {pages} {405} (\bibinfo {year}
  {2010})},\ \Eprint {https://arxiv.org/abs/1002.0329} {arXiv:1002.0329
  [hep-ph]} \BibitemShut {NoStop}%
\bibitem [{\citenamefont {Moriyama}\ \emph {et~al.}(1998)\citenamefont
  {Moriyama} \emph {et~al.}}]{Sumico}%
  \BibitemOpen
  \bibfield  {author} {\bibinfo {author} {\bibfnamefont {S.}~\bibnamefont
  {Moriyama}} \emph {et~al.} (\bibinfo {collaboration} {SUMICO
  Collaboration}),\ }\href
  {https://doi.org/https://doi.org/10.1016/S0370-2693(98)00766-7} {\bibfield
  {journal} {\bibinfo  {journal} {Physics Letters B}\ }\textbf {\bibinfo
  {volume} {434}},\ \bibinfo {pages} {147} (\bibinfo {year}
  {1998})}\BibitemShut {NoStop}%
\bibitem [{\citenamefont {Altenm\"uller}\ \emph {et~al.}(2024)\citenamefont
  {Altenm\"uller} \emph {et~al.}}]{PhysRevLett.133.221005}%
  \BibitemOpen
  \bibfield  {author} {\bibinfo {author} {\bibfnamefont {K.}~\bibnamefont
  {Altenm\"uller}} \emph {et~al.} (\bibinfo {collaboration} {CAST
  Collaboration}),\ }\href {https://doi.org/10.1103/PhysRevLett.133.221005}
  {\bibfield  {journal} {\bibinfo  {journal} {Phys. Rev. Lett.}\ }\textbf
  {\bibinfo {volume} {133}},\ \bibinfo {pages} {221005} (\bibinfo {year}
  {2024})}\BibitemShut {NoStop}%
\bibitem [{\citenamefont {Asztalos}\ \emph {et~al.}(2010)\citenamefont
  {Asztalos} \emph {et~al.}}]{Asztalos_2010}%
  \BibitemOpen
  \bibfield  {author} {\bibinfo {author} {\bibfnamefont {S.~J.}\ \bibnamefont
  {Asztalos}} \emph {et~al.},\ }\bibfield  {journal} {\bibinfo  {journal}
  {Phys. Rev. Lett.}\ }\textbf {\bibinfo {volume} {104}},\ \href
  {https://doi.org/10.1103/physrevlett.104.041301}
  {10.1103/physrevlett.104.041301} (\bibinfo {year} {2010})\BibitemShut
  {NoStop}%
\bibitem [{\citenamefont {Du}\ \emph {et~al.}(2018)\citenamefont {Du} \emph
  {et~al.}}]{Du_2018}%
  \BibitemOpen
  \bibfield  {author} {\bibinfo {author} {\bibfnamefont {N.}~\bibnamefont {Du}}
  \emph {et~al.},\ }\bibfield  {journal} {\bibinfo  {journal} {Phys. Rev.
  Lett.}\ }\textbf {\bibinfo {volume} {120}},\ \href
  {https://doi.org/10.1103/physrevlett.120.151301}
  {10.1103/physrevlett.120.151301} (\bibinfo {year} {2018})\BibitemShut
  {NoStop}%
\bibitem [{\citenamefont {Braine}\ \emph {et~al.}(2020)\citenamefont {Braine}
  \emph {et~al.}}]{Braine_2020}%
  \BibitemOpen
  \bibfield  {author} {\bibinfo {author} {\bibfnamefont {T.}~\bibnamefont
  {Braine}} \emph {et~al.},\ }\bibfield  {journal} {\bibinfo  {journal} {Phys.
  Rev. Lett.}\ }\textbf {\bibinfo {volume} {124}},\ \href
  {https://doi.org/10.1103/physrevlett.124.101303}
  {10.1103/physrevlett.124.101303} (\bibinfo {year} {2020})\BibitemShut
  {NoStop}%
\bibitem [{\citenamefont {Bartram}\ \emph {et~al.}(2021)\citenamefont {Bartram}
  \emph {et~al.}}]{Bartram_2021}%
  \BibitemOpen
  \bibfield  {author} {\bibinfo {author} {\bibfnamefont {C.}~\bibnamefont
  {Bartram}} \emph {et~al.} (\bibinfo {collaboration} {ADMX}),\ }\href
  {https://doi.org/10.1103/PhysRevLett.127.261803} {\bibfield  {journal}
  {\bibinfo  {journal} {Phys. Rev. Lett.}\ }\textbf {\bibinfo {volume} {127}},\
  \bibinfo {pages} {261803} (\bibinfo {year} {2021})},\ \Eprint
  {https://arxiv.org/abs/2110.06096} {arXiv:2110.06096 [hep-ex]} \BibitemShut
  {NoStop}%
\bibitem [{\citenamefont {Boutan}\ \emph {et~al.}(2018)\citenamefont {Boutan}
  \emph {et~al.}}]{Boutan_2018}%
  \BibitemOpen
  \bibfield  {author} {\bibinfo {author} {\bibfnamefont {C.}~\bibnamefont
  {Boutan}} \emph {et~al.},\ }\bibfield  {journal} {\bibinfo  {journal} {Phys.
  Rev. Lett.}\ }\textbf {\bibinfo {volume} {121}},\ \href
  {https://doi.org/10.1103/physrevlett.121.261302}
  {10.1103/physrevlett.121.261302} (\bibinfo {year} {2018})\BibitemShut
  {NoStop}%
\bibitem [{\citenamefont {Bartram}\ \emph {et~al.}(2023)\citenamefont {Bartram}
  \emph {et~al.}}]{Bartram_2023}%
  \BibitemOpen
  \bibfield  {author} {\bibinfo {author} {\bibfnamefont {C.}~\bibnamefont
  {Bartram}} \emph {et~al.},\ }\bibfield  {journal} {\bibinfo  {journal}
  {Review of Scientific Instruments}\ }\textbf {\bibinfo {volume} {94}},\ \href
  {https://doi.org/10.1063/5.0122907} {10.1063/5.0122907} (\bibinfo {year}
  {2023})\BibitemShut {NoStop}%
\bibitem [{\citenamefont {Hagmann}\ \emph {et~al.}(1990)\citenamefont {Hagmann}
  \emph {et~al.}}]{PhysRevD.42.1297}%
  \BibitemOpen
  \bibfield  {author} {\bibinfo {author} {\bibfnamefont {C.}~\bibnamefont
  {Hagmann}} \emph {et~al.},\ }\href {https://doi.org/10.1103/PhysRevD.42.1297}
  {\bibfield  {journal} {\bibinfo  {journal} {Phys. Rev. D}\ }\textbf {\bibinfo
  {volume} {42}},\ \bibinfo {pages} {1297} (\bibinfo {year}
  {1990})}\BibitemShut {NoStop}%
\bibitem [{\citenamefont {Zhong}\ \emph {et~al.}(2018)\citenamefont {Zhong}
  \emph {et~al.}}]{Zhong_2018}%
  \BibitemOpen
  \bibfield  {author} {\bibinfo {author} {\bibfnamefont {L.}~\bibnamefont
  {Zhong}} \emph {et~al.},\ }\bibfield  {journal} {\bibinfo  {journal} {Phys.
  Rev. D}\ }\textbf {\bibinfo {volume} {97}},\ \href
  {https://doi.org/10.1103/physrevd.97.092001} {10.1103/physrevd.97.092001}
  (\bibinfo {year} {2018})\BibitemShut {NoStop}%
\bibitem [{\citenamefont {Backes}\ \emph {et~al.}(2021)\citenamefont {Backes}
  \emph {et~al.}}]{Backes_2021}%
  \BibitemOpen
  \bibfield  {author} {\bibinfo {author} {\bibfnamefont {K.~M.}\ \bibnamefont
  {Backes}} \emph {et~al.},\ }\href
  {https://doi.org/10.1038/s41586-021-03226-7} {\bibfield  {journal} {\bibinfo
  {journal} {Nature}\ }\textbf {\bibinfo {volume} {590}},\ \bibinfo {pages}
  {238–242} (\bibinfo {year} {2021})}\BibitemShut {NoStop}%
\bibitem [{\citenamefont {Jewell}\ \emph {et~al.}(2023)\citenamefont {Jewell}
  \emph {et~al.}}]{haystaccollaboration2023new}%
  \BibitemOpen
  \bibfield  {author} {\bibinfo {author} {\bibfnamefont {M.~J.}\ \bibnamefont
  {Jewell}} \emph {et~al.} (\bibinfo {collaboration} {HAYSTAC Collaboration}),\
  }\href {https://doi.org/10.1103/PhysRevD.107.072007} {\bibfield  {journal}
  {\bibinfo  {journal} {Phys. Rev. D}\ }\textbf {\bibinfo {volume} {107}},\
  \bibinfo {pages} {072007} (\bibinfo {year} {2023})}\BibitemShut {NoStop}%
\bibitem [{\citenamefont {Lee}\ \emph {et~al.}(2020)\citenamefont {Lee} \emph
  {et~al.}}]{Lee_2020}%
  \BibitemOpen
  \bibfield  {author} {\bibinfo {author} {\bibfnamefont {S.}~\bibnamefont
  {Lee}} \emph {et~al.},\ }\href
  {https://doi.org/10.1103/PhysRevLett.124.101802} {\bibfield  {journal}
  {\bibinfo  {journal} {Phys. Rev. Lett.}\ }\textbf {\bibinfo {volume} {124}},\
  \bibinfo {pages} {101802} (\bibinfo {year} {2020})},\ \Eprint
  {https://arxiv.org/abs/2001.05102} {arXiv:2001.05102 [hep-ex]} \BibitemShut
  {NoStop}%
\bibitem [{\citenamefont {Jeong}\ \emph {et~al.}(2020)\citenamefont {Jeong}
  \emph {et~al.}}]{Jeong_2020}%
  \BibitemOpen
  \bibfield  {author} {\bibinfo {author} {\bibfnamefont {J.}~\bibnamefont
  {Jeong}} \emph {et~al.},\ }\bibfield  {journal} {\bibinfo  {journal} {Phys.
  Rev. Lett.}\ }\textbf {\bibinfo {volume} {125}},\ \href
  {https://doi.org/10.1103/physrevlett.125.221302}
  {10.1103/physrevlett.125.221302} (\bibinfo {year} {2020})\BibitemShut
  {NoStop}%
\bibitem [{\citenamefont {Kwon}\ \emph {et~al.}(2021)\citenamefont {Kwon} \emph
  {et~al.}}]{Kwon_2021}%
  \BibitemOpen
  \bibfield  {author} {\bibinfo {author} {\bibfnamefont {O.}~\bibnamefont
  {Kwon}} \emph {et~al.} (\bibinfo {collaboration} {CAPP}),\ }\href
  {https://doi.org/10.1103/PhysRevLett.126.191802} {\bibfield  {journal}
  {\bibinfo  {journal} {Phys. Rev. Lett.}\ }\textbf {\bibinfo {volume} {126}},\
  \bibinfo {pages} {191802} (\bibinfo {year} {2021})},\ \Eprint
  {https://arxiv.org/abs/2012.10764} {arXiv:2012.10764 [hep-ex]} \BibitemShut
  {NoStop}%
\bibitem [{\citenamefont {Lee}\ \emph {et~al.}(2022)\citenamefont {Lee} \emph
  {et~al.}}]{Lee_2022}%
  \BibitemOpen
  \bibfield  {author} {\bibinfo {author} {\bibfnamefont {Y.}~\bibnamefont
  {Lee}} \emph {et~al.},\ }\bibfield  {journal} {\bibinfo  {journal} {Phys.
  Rev. Lett.}\ }\textbf {\bibinfo {volume} {128}},\ \href
  {https://doi.org/10.1103/physrevlett.128.241805}
  {10.1103/physrevlett.128.241805} (\bibinfo {year} {2022})\BibitemShut
  {NoStop}%
\bibitem [{\citenamefont {Kim}\ \emph {et~al.}(2023)\citenamefont {Kim} \emph
  {et~al.}}]{Kim_2023}%
  \BibitemOpen
  \bibfield  {author} {\bibinfo {author} {\bibfnamefont {J.}~\bibnamefont
  {Kim}} \emph {et~al.},\ }\href
  {https://doi.org/10.1103/PhysRevLett.130.091602} {\bibfield  {journal}
  {\bibinfo  {journal} {Phys. Rev. Lett.}\ }\textbf {\bibinfo {volume} {130}},\
  \bibinfo {pages} {091602} (\bibinfo {year} {2023})},\ \Eprint
  {https://arxiv.org/abs/2207.13597} {arXiv:2207.13597 [hep-ex]} \BibitemShut
  {NoStop}%
\bibitem [{\citenamefont {Yi}\ \emph {et~al.}(2023)\citenamefont {Yi} \emph
  {et~al.}}]{Yi_2023}%
  \BibitemOpen
  \bibfield  {author} {\bibinfo {author} {\bibfnamefont {A.~K.}\ \bibnamefont
  {Yi}} \emph {et~al.},\ }\href
  {https://doi.org/10.1103/PhysRevLett.130.071002} {\bibfield  {journal}
  {\bibinfo  {journal} {Phys. Rev. Lett.}\ }\textbf {\bibinfo {volume} {130}},\
  \bibinfo {pages} {071002} (\bibinfo {year} {2023})},\ \Eprint
  {https://arxiv.org/abs/2210.10961} {arXiv:2210.10961 [hep-ex]} \BibitemShut
  {NoStop}%
\bibitem [{\citenamefont {Yang}\ \emph {et~al.}(2023)\citenamefont {Yang} \emph
  {et~al.}}]{Yang_2023}%
  \BibitemOpen
  \bibfield  {author} {\bibinfo {author} {\bibfnamefont {B.}~\bibnamefont
  {Yang}} \emph {et~al.},\ }\bibfield  {journal} {\bibinfo  {journal} {Phys.
  Rev. Lett.}\ }\textbf {\bibinfo {volume} {131}},\ \href
  {https://doi.org/10.1103/physrevlett.131.081801}
  {10.1103/physrevlett.131.081801} (\bibinfo {year} {2023})\BibitemShut
  {NoStop}%
\bibitem [{\citenamefont {Kim}\ \emph {et~al.}(2024)\citenamefont {Kim} \emph
  {et~al.}}]{kim2023experimental}%
  \BibitemOpen
  \bibfield  {author} {\bibinfo {author} {\bibfnamefont {Y.}~\bibnamefont
  {Kim}} \emph {et~al.},\ }\href
  {https://doi.org/10.1103/PhysRevLett.133.051802} {\bibfield  {journal}
  {\bibinfo  {journal} {Phys. Rev. Lett.}\ }\textbf {\bibinfo {volume} {133}},\
  \bibinfo {pages} {051802} (\bibinfo {year} {2024})}\BibitemShut {NoStop}%
\bibitem [{\citenamefont {Ahn}\ \emph {et~al.}(2024)\citenamefont {Ahn} \emph
  {et~al.}}]{PhysRevX.14.031023}%
  \BibitemOpen
  \bibfield  {author} {\bibinfo {author} {\bibfnamefont {S.}~\bibnamefont
  {Ahn}} \emph {et~al.},\ }\href {https://doi.org/10.1103/PhysRevX.14.031023}
  {\bibfield  {journal} {\bibinfo  {journal} {Phys. Rev. X}\ }\textbf {\bibinfo
  {volume} {14}},\ \bibinfo {pages} {031023} (\bibinfo {year}
  {2024})}\BibitemShut {NoStop}%
\bibitem [{\citenamefont {Quiskamp}\ \emph {et~al.}(2022)\citenamefont
  {Quiskamp} \emph {et~al.}}]{Quiskamp_2022}%
  \BibitemOpen
  \bibfield  {author} {\bibinfo {author} {\bibfnamefont {A.}~\bibnamefont
  {Quiskamp}} \emph {et~al.},\ }\bibfield  {journal} {\bibinfo  {journal}
  {Science Advances}\ }\textbf {\bibinfo {volume} {8}},\ \href
  {https://doi.org/10.1126/sciadv.abq3765} {10.1126/sciadv.abq3765} (\bibinfo
  {year} {2022})\BibitemShut {NoStop}%
\bibitem [{\citenamefont {Quiskamp}\ \emph {et~al.}(2024)\citenamefont
  {Quiskamp} \emph {et~al.}}]{Abeln:2021}%
  \BibitemOpen
  \bibfield  {author} {\bibinfo {author} {\bibfnamefont {A.}~\bibnamefont
  {Quiskamp}} \emph {et~al.},\ }\href
  {https://doi.org/10.1103/PhysRevLett.132.031601} {\bibfield  {journal}
  {\bibinfo  {journal} {Phys. Rev. Lett.}\ }\textbf {\bibinfo {volume} {132}},\
  \bibinfo {pages} {031601} (\bibinfo {year} {2024})}\BibitemShut {NoStop}%
\bibitem [{\citenamefont {McAllister}\ \emph {et~al.}(2017)\citenamefont
  {McAllister} \emph {et~al.}}]{McAllister_2017}%
  \BibitemOpen
  \bibfield  {author} {\bibinfo {author} {\bibfnamefont {B.~T.}\ \bibnamefont
  {McAllister}} \emph {et~al.},\ }\href
  {https://doi.org/10.1016/j.dark.2017.09.010} {\bibfield  {journal} {\bibinfo
  {journal} {Physics of the Dark Universe}\ }\textbf {\bibinfo {volume} {18}},\
  \bibinfo {pages} {67–72} (\bibinfo {year} {2017})}\BibitemShut {NoStop}%
\bibitem [{\citenamefont {Alesini}\ \emph {et~al.}(2019)\citenamefont {Alesini}
  \emph {et~al.}}]{Alesini_2019}%
  \BibitemOpen
  \bibfield  {author} {\bibinfo {author} {\bibfnamefont {D.}~\bibnamefont
  {Alesini}} \emph {et~al.},\ }\bibfield  {journal} {\bibinfo  {journal} {Phys.
  Rev. D}\ }\textbf {\bibinfo {volume} {99}},\ \href
  {https://doi.org/10.1103/physrevd.99.101101} {10.1103/physrevd.99.101101}
  (\bibinfo {year} {2019})\BibitemShut {NoStop}%
\bibitem [{\citenamefont {Alesini}\ \emph {et~al.}(2021)\citenamefont {Alesini}
  \emph {et~al.}}]{Alesini_2021}%
  \BibitemOpen
  \bibfield  {author} {\bibinfo {author} {\bibfnamefont {D.}~\bibnamefont
  {Alesini}} \emph {et~al.},\ }\bibfield  {journal} {\bibinfo  {journal} {Phys.
  Rev. D}\ }\textbf {\bibinfo {volume} {103}},\ \href
  {https://doi.org/10.1103/physrevd.103.102004} {10.1103/physrevd.103.102004}
  (\bibinfo {year} {2021})\BibitemShut {NoStop}%
\bibitem [{\citenamefont {Di~Vora}\ \emph {et~al.}(2023)\citenamefont {Di~Vora}
  \emph {et~al.}}]{Di_Vora_2023}%
  \BibitemOpen
  \bibfield  {author} {\bibinfo {author} {\bibfnamefont {R.}~\bibnamefont
  {Di~Vora}} \emph {et~al.},\ }\bibfield  {journal} {\bibinfo  {journal} {Phys.
  Rev. D}\ }\textbf {\bibinfo {volume} {108}},\ \href
  {https://doi.org/10.1103/physrevd.108.062005} {10.1103/physrevd.108.062005}
  (\bibinfo {year} {2023})\BibitemShut {NoStop}%
\bibitem [{\citenamefont {Devlin}\ \emph {et~al.}(2021)\citenamefont {Devlin}
  \emph {et~al.}}]{Devlin:2021fpq}%
  \BibitemOpen
  \bibfield  {author} {\bibinfo {author} {\bibfnamefont {J.~A.}\ \bibnamefont
  {Devlin}} \emph {et~al.},\ }\href
  {https://doi.org/10.1103/PhysRevLett.126.041301} {\bibfield  {journal}
  {\bibinfo  {journal} {Phys. Rev. Lett.}\ }\textbf {\bibinfo {volume} {126}},\
  \bibinfo {pages} {041301} (\bibinfo {year} {2021})},\ \Eprint
  {https://arxiv.org/abs/2101.11290} {arXiv:2101.11290 [astro-ph.CO]}
  \BibitemShut {NoStop}%
\bibitem [{\citenamefont {Ouellet}\ \emph {et~al.}(2019)\citenamefont {Ouellet}
  \emph {et~al.}}]{Ouellet_2019}%
  \BibitemOpen
  \bibfield  {author} {\bibinfo {author} {\bibfnamefont {J.~L.}\ \bibnamefont
  {Ouellet}} \emph {et~al.},\ }\bibfield  {journal} {\bibinfo  {journal} {Phys.
  Rev. Lett.}\ }\textbf {\bibinfo {volume} {122}},\ \href
  {https://doi.org/10.1103/physrevlett.122.121802}
  {10.1103/physrevlett.122.121802} (\bibinfo {year} {2019})\BibitemShut
  {NoStop}%
\bibitem [{\citenamefont {Salemi}\ \emph {et~al.}(2021)\citenamefont {Salemi}
  \emph {et~al.}}]{Salemi_2021}%
  \BibitemOpen
  \bibfield  {author} {\bibinfo {author} {\bibfnamefont {C.~P.}\ \bibnamefont
  {Salemi}} \emph {et~al.},\ }\bibfield  {journal} {\bibinfo  {journal} {Phys.
  Rev. Lett.}\ }\textbf {\bibinfo {volume} {127}},\ \href
  {https://doi.org/10.1103/physrevlett.127.081801}
  {10.1103/physrevlett.127.081801} (\bibinfo {year} {2021})\BibitemShut
  {NoStop}%
\bibitem [{\citenamefont {Crisosto}\ \emph {et~al.}(2020)\citenamefont
  {Crisosto} \emph {et~al.}}]{Crisosto:2019fcj}%
  \BibitemOpen
  \bibfield  {author} {\bibinfo {author} {\bibfnamefont {N.}~\bibnamefont
  {Crisosto}} \emph {et~al.},\ }\href
  {https://doi.org/10.1103/PhysRevLett.124.241101} {\bibfield  {journal}
  {\bibinfo  {journal} {Phys. Rev. Lett.}\ }\textbf {\bibinfo {volume} {124}},\
  \bibinfo {pages} {241101} (\bibinfo {year} {2020})},\ \Eprint
  {https://arxiv.org/abs/1911.05772} {arXiv:1911.05772 [astro-ph.CO]}
  \BibitemShut {NoStop}%
\bibitem [{\citenamefont {B{\"a}hre}\ \emph {et~al.}(2013)\citenamefont
  {B{\"a}hre} \emph {et~al.}}]{RBahre_2013}%
  \BibitemOpen
  \bibfield  {author} {\bibinfo {author} {\bibfnamefont {R.}~\bibnamefont
  {B{\"a}hre}} \emph {et~al.},\ }\href
  {https://doi.org/10.1088/1748-0221/8/09/T09001} {\bibfield  {journal}
  {\bibinfo  {journal} {Journal of Instrumentation}\ }\textbf {\bibinfo
  {volume} {8}}\bibinfo  {number} { (09)},\ \bibinfo {pages}
  {T09001}}\BibitemShut {NoStop}%
\bibitem [{\citenamefont {Armengaud}\ \emph {et~al.}(2019)\citenamefont
  {Armengaud} \emph {et~al.}}]{PhysPotIAXO}%
  \BibitemOpen
\bibfield  {number} {  }\bibfield  {author} {\bibinfo {author} {\bibfnamefont
  {E.}~\bibnamefont {Armengaud}} \emph {et~al.} (\bibinfo {collaboration} {IAXO
  Collaboration}),\ }\href {https://doi.org/10.1088/1475-7516/2019/06/047}
  {\bibfield  {journal} {\bibinfo  {journal} {Journal of Cosmology and
  Astroparticle Physics}\ }\textbf {\bibinfo {volume} {2019}}\bibinfo  {number}
  { (06)},\ \bibinfo {pages} {047}}\BibitemShut {NoStop}%
\bibitem [{\citenamefont {Armengaud}\ \emph {et~al.}(2014)\citenamefont
  {Armengaud} \emph {et~al.}}]{Armengaud:2014gea}%
  \BibitemOpen
\bibfield  {number} {  }\bibfield  {author} {\bibinfo {author} {\bibfnamefont
  {E.}~\bibnamefont {Armengaud}} \emph {et~al.},\ }\href
  {https://doi.org/10.1088/1748-0221/9/05/T05002} {\bibfield  {journal}
  {\bibinfo  {journal} {Journal of Instrumentation}\ }\textbf {\bibinfo
  {volume} {9}}\bibinfo  {number} { (05)},\ \bibinfo {pages}
  {T05002}}\BibitemShut {NoStop}%
\bibitem [{\citenamefont {Sikivie}(1983)}]{Sikivie:1983ip}%
  \BibitemOpen
\bibfield  {number} {  }\bibfield  {author} {\bibinfo {author} {\bibfnamefont
  {P.}~\bibnamefont {Sikivie}},\ }\href
  {https://doi.org/10.1103/PhysRevLett.51.1415, 10.1103/PhysRevLett.52.695.2}
  {\bibfield  {journal} {\bibinfo  {journal} {Phys. Rev. Lett.}\ }\textbf
  {\bibinfo {volume} {51}},\ \bibinfo {pages} {1415} (\bibinfo {year}
  {1983})}\BibitemShut {NoStop}%
\bibitem [{\citenamefont {{Tsuneta}}\ \emph {et~al.}(1991)\citenamefont
  {{Tsuneta}} \emph {et~al.}}]{1991SoPh..136...37T}%
  \BibitemOpen
  \bibfield  {author} {\bibinfo {author} {\bibfnamefont {S.}~\bibnamefont
  {{Tsuneta}}} \emph {et~al.},\ }\href {https://doi.org/10.1007/BF00151694}
  {\bibfield  {journal} {\bibinfo  {journal} {Solar Physics}\ }\textbf
  {\bibinfo {volume} {136}},\ \bibinfo {pages} {37} (\bibinfo {year}
  {1991})}\BibitemShut {NoStop}%
\bibitem [{\citenamefont {Carlson}\ and\ \citenamefont
  {Tseng}(1996)}]{Carlson:1995xf}%
  \BibitemOpen
  \bibfield  {author} {\bibinfo {author} {\bibfnamefont {E.~D.}\ \bibnamefont
  {Carlson}}\ and\ \bibinfo {author} {\bibfnamefont {L.-S.}\ \bibnamefont
  {Tseng}},\ }\href {https://doi.org/10.1016/0370-2693(95)01250-8} {\bibfield
  {journal} {\bibinfo  {journal} {Phys. Lett. B}\ }\textbf {\bibinfo {volume}
  {365}},\ \bibinfo {pages} {193} (\bibinfo {year} {1996})}\BibitemShut
  {NoStop}%
\bibitem [{\citenamefont {{Hannah}}\ \emph {et~al.}(2007)\citenamefont
  {{Hannah}} \emph {et~al.}}]{2007ApJ...659L..77H}%
  \BibitemOpen
  \bibfield  {author} {\bibinfo {author} {\bibfnamefont {I.~G.}\ \bibnamefont
  {{Hannah}}} \emph {et~al.},\ }\href {https://doi.org/10.1086/516750}
  {\bibfield  {journal} {\bibinfo  {journal} {\apjl}\ }\textbf {\bibinfo
  {volume} {659}},\ \bibinfo {pages} {L77} (\bibinfo {year} {2007})},\ \Eprint
  {https://arxiv.org/abs/astro-ph/0702726} {astro-ph/0702726} \BibitemShut
  {NoStop}%
\bibitem [{\citenamefont {{Hannah}}\ \emph {et~al.}(2010)\citenamefont
  {{Hannah}}, \citenamefont {{Hudson}}, \citenamefont {{Hurford}},\ and\
  \citenamefont {{Lin}}}]{2010ApJ...724..487H}%
  \BibitemOpen
  \bibfield  {author} {\bibinfo {author} {\bibfnamefont {I.~G.}\ \bibnamefont
  {{Hannah}}}, \bibinfo {author} {\bibfnamefont {H.~S.}\ \bibnamefont
  {{Hudson}}}, \bibinfo {author} {\bibfnamefont {G.~J.}\ \bibnamefont
  {{Hurford}}},\ and\ \bibinfo {author} {\bibfnamefont {R.~P.}\ \bibnamefont
  {{Lin}}},\ }\href {https://doi.org/10.1088/0004-637X/724/1/487} {\bibfield
  {journal} {\bibinfo  {journal} {\apj}\ }\textbf {\bibinfo {volume} {724}},\
  \bibinfo {pages} {487} (\bibinfo {year} {2010})},\ \Eprint
  {https://arxiv.org/abs/1009.2918} {arXiv:1009.2918 [astro-ph.SR]}
  \BibitemShut {NoStop}%
\bibitem [{\citenamefont {{Hudson}}\ \emph {et~al.}(2012)\citenamefont
  {{Hudson}} \emph {et~al.}}]{2012ASPC..455...25H}%
  \BibitemOpen
  \bibfield  {author} {\bibinfo {author} {\bibfnamefont {H.~S.}\ \bibnamefont
  {{Hudson}}} \emph {et~al.},\ }in\ \href
  {https://doi.org/10.48550/arXiv.1201.4607} {\emph {\bibinfo {booktitle} {4th
  Hinode Science Meeting: Unsolved Problems and Recent Insights}}},\ \bibinfo
  {series} {Astronomical Society of the Pacific Conference Series}, Vol.\
  \bibinfo {volume} {455}\ (\bibinfo {year} {2012})\ p.~\bibinfo {pages}
  {25}\BibitemShut {NoStop}%
\bibitem [{\citenamefont {Harrison}\ \emph {et~al.}(2013)\citenamefont
  {Harrison} \emph {et~al.}}]{2013ApJ...770..103H}%
  \BibitemOpen
  \bibfield  {author} {\bibinfo {author} {\bibfnamefont {F.~A.}\ \bibnamefont
  {Harrison}} \emph {et~al.},\ }\href
  {https://doi.org/10.1088/0004-637X/770/2/103} {\bibfield  {journal} {\bibinfo
   {journal} {The Astrophysical Journal}\ }\textbf {\bibinfo {volume} {770}},\
  \bibinfo {pages} {103} (\bibinfo {year} {2013})}\BibitemShut {NoStop}%
\bibitem [{\citenamefont {Li}\ \emph {et~al.}(2021)\citenamefont {Li} \emph
  {et~al.}}]{Li:2024zst}%
  \BibitemOpen
  \bibfield  {author} {\bibinfo {author} {\bibfnamefont {H.-J.}\ \bibnamefont
  {Li}} \emph {et~al.},\ }\href {https://doi.org/10.1103/PhysRevD.103.083003}
  {\bibfield  {journal} {\bibinfo  {journal} {Phys. Rev. D}\ }\textbf {\bibinfo
  {volume} {103}},\ \bibinfo {pages} {083003} (\bibinfo {year}
  {2021})}\BibitemShut {NoStop}%
\bibitem [{\citenamefont {Li}\ \emph {et~al.}(2022)\citenamefont {Li},
  \citenamefont {Bi},\ and\ \citenamefont {Yin}}]{Li_2022}%
  \BibitemOpen
  \bibfield  {author} {\bibinfo {author} {\bibfnamefont {H.-J.}\ \bibnamefont
  {Li}}, \bibinfo {author} {\bibfnamefont {X.-J.}\ \bibnamefont {Bi}},\ and\
  \bibinfo {author} {\bibfnamefont {P.-F.}\ \bibnamefont {Yin}},\ }\href
  {https://doi.org/10.1088/1674-1137/ac6d4f} {\bibfield  {journal} {\bibinfo
  {journal} {Chinese Physics C}\ }\textbf {\bibinfo {volume} {46}},\ \bibinfo
  {pages} {085105} (\bibinfo {year} {2022})}\BibitemShut {NoStop}%
\bibitem [{\citenamefont {Reynolds}\ \emph {et~al.}(2020)\citenamefont
  {Reynolds} \emph {et~al.}}]{Reynolds_2020}%
  \BibitemOpen
  \bibfield  {author} {\bibinfo {author} {\bibfnamefont {C.~S.}\ \bibnamefont
  {Reynolds}} \emph {et~al.},\ }\href
  {https://doi.org/10.3847/1538-4357/ab6a0c} {\bibfield  {journal} {\bibinfo
  {journal} {The Astrophysical Journal}\ }\textbf {\bibinfo {volume} {890}},\
  \bibinfo {pages} {59} (\bibinfo {year} {2020})}\BibitemShut {NoStop}%
\bibitem [{\citenamefont {Abramowski}\ \emph {et~al.}(2013)\citenamefont
  {Abramowski} \emph {et~al.}}]{Abramowski_2013}%
  \BibitemOpen
  \bibfield  {author} {\bibinfo {author} {\bibfnamefont {A.}~\bibnamefont
  {Abramowski}} \emph {et~al.},\ }\bibfield  {journal} {\bibinfo  {journal}
  {Phys. Rev. D}\ }\textbf {\bibinfo {volume} {88}},\ \href
  {https://doi.org/10.1103/physrevd.88.102003} {10.1103/physrevd.88.102003}
  (\bibinfo {year} {2013})\BibitemShut {NoStop}%
\bibitem [{\citenamefont {Ajello}\ \emph {et~al.}(2016)\citenamefont {Ajello}
  \emph {et~al.}}]{Ajello_2016}%
  \BibitemOpen
  \bibfield  {author} {\bibinfo {author} {\bibfnamefont {M.}~\bibnamefont
  {Ajello}} \emph {et~al.},\ }\bibfield  {journal} {\bibinfo  {journal} {Phys.
  Rev. Lett.}\ }\textbf {\bibinfo {volume} {116}},\ \href
  {https://doi.org/10.1103/physrevlett.116.161101}
  {10.1103/physrevlett.116.161101} (\bibinfo {year} {2016})\BibitemShut
  {NoStop}%
\bibitem [{\citenamefont {Davies}\ \emph {et~al.}(2023)\citenamefont {Davies},
  \citenamefont {Meyer},\ and\ \citenamefont {Cotter}}]{Davies_2023}%
  \BibitemOpen
  \bibfield  {author} {\bibinfo {author} {\bibfnamefont {J.}~\bibnamefont
  {Davies}}, \bibinfo {author} {\bibfnamefont {M.}~\bibnamefont {Meyer}},\ and\
  \bibinfo {author} {\bibfnamefont {G.}~\bibnamefont {Cotter}},\ }\bibfield
  {journal} {\bibinfo  {journal} {Phys. Rev. D}\ }\textbf {\bibinfo {volume}
  {107}},\ \href {https://doi.org/10.1103/physrevd.107.083027}
  {10.1103/physrevd.107.083027} (\bibinfo {year} {2023})\BibitemShut {NoStop}%
\bibitem [{\citenamefont {Jacobsen}\ \emph {et~al.}(2023)\citenamefont
  {Jacobsen}, \citenamefont {Linden},\ and\ \citenamefont
  {Freese}}]{Jacobsen:2022swa}%
  \BibitemOpen
  \bibfield  {author} {\bibinfo {author} {\bibfnamefont {S.}~\bibnamefont
  {Jacobsen}}, \bibinfo {author} {\bibfnamefont {T.}~\bibnamefont {Linden}},\
  and\ \bibinfo {author} {\bibfnamefont {K.}~\bibnamefont {Freese}},\ }\href
  {https://doi.org/10.1088/1475-7516/2023/10/009} {\bibfield  {journal}
  {\bibinfo  {journal} {Journal of Cosmology and Astroparticle Physics}\
  }\textbf {\bibinfo {volume} {2023}}\bibinfo  {number} { (10)},\ \bibinfo
  {pages} {009}}\BibitemShut {NoStop}%
\bibitem [{\citenamefont {Xiao}\ \emph {et~al.}(2021)\citenamefont {Xiao} \emph
  {et~al.}}]{Xiao:2020pra}%
  \BibitemOpen
\bibfield  {number} {  }\bibfield  {author} {\bibinfo {author} {\bibfnamefont
  {M.}~\bibnamefont {Xiao}} \emph {et~al.},\ }\href
  {https://doi.org/10.1103/PhysRevLett.126.031101} {\bibfield  {journal}
  {\bibinfo  {journal} {Phys. Rev. Lett.}\ }\textbf {\bibinfo {volume} {126}},\
  \bibinfo {pages} {031101} (\bibinfo {year} {2021})},\ \Eprint
  {https://arxiv.org/abs/2009.09059} {arXiv:2009.09059 [astro-ph.HE]}
  \BibitemShut {NoStop}%
\bibitem [{\citenamefont {Ning}\ and\ \citenamefont
  {Safdi}(2025)}]{Ning:2024ek}%
  \BibitemOpen
  \bibfield  {author} {\bibinfo {author} {\bibfnamefont {O.}~\bibnamefont
  {Ning}}\ and\ \bibinfo {author} {\bibfnamefont {B.~R.}\ \bibnamefont
  {Safdi}},\ }\href {https://doi.org/10.1103/PhysRevLett.134.171003} {\bibfield
   {journal} {\bibinfo  {journal} {Phys. Rev. Lett.}\ }\textbf {\bibinfo
  {volume} {134}},\ \bibinfo {pages} {171003} (\bibinfo {year}
  {2025})}\BibitemShut {NoStop}%
\bibitem [{\citenamefont {Dessert}\ \emph
  {et~al.}(2022{\natexlab{a}})\citenamefont {Dessert}, \citenamefont {Dunsky},\
  and\ \citenamefont {Safdi}}]{Dessert_2022_1}%
  \BibitemOpen
  \bibfield  {author} {\bibinfo {author} {\bibfnamefont {C.}~\bibnamefont
  {Dessert}}, \bibinfo {author} {\bibfnamefont {D.}~\bibnamefont {Dunsky}},\
  and\ \bibinfo {author} {\bibfnamefont {B.~R.}\ \bibnamefont {Safdi}},\
  }\bibfield  {journal} {\bibinfo  {journal} {Phys. Rev. D}\ }\textbf {\bibinfo
  {volume} {105}},\ \href {https://doi.org/10.1103/physrevd.105.103034}
  {10.1103/physrevd.105.103034} (\bibinfo {year}
  {2022}{\natexlab{a}})\BibitemShut {NoStop}%
\bibitem [{\citenamefont {Dessert}\ \emph
  {et~al.}(2022{\natexlab{b}})\citenamefont {Dessert}, \citenamefont {Long},\
  and\ \citenamefont {Safdi}}]{Dessert_2022_2}%
  \BibitemOpen
  \bibfield  {author} {\bibinfo {author} {\bibfnamefont {C.}~\bibnamefont
  {Dessert}}, \bibinfo {author} {\bibfnamefont {A.~J.}\ \bibnamefont {Long}},\
  and\ \bibinfo {author} {\bibfnamefont {B.~R.}\ \bibnamefont {Safdi}},\
  }\bibfield  {journal} {\bibinfo  {journal} {Phys. Rev. Lett.}\ }\textbf
  {\bibinfo {volume} {128}},\ \href
  {https://doi.org/10.1103/physrevlett.128.071102}
  {10.1103/physrevlett.128.071102} (\bibinfo {year}
  {2022}{\natexlab{b}})\BibitemShut {NoStop}%
\bibitem [{\citenamefont {Noordhuis}\ \emph {et~al.}(2023)\citenamefont
  {Noordhuis} \emph {et~al.}}]{Noordhuis:2022ljw}%
  \BibitemOpen
  \bibfield  {author} {\bibinfo {author} {\bibfnamefont {D.}~\bibnamefont
  {Noordhuis}} \emph {et~al.},\ }\href
  {https://doi.org/10.1103/PhysRevLett.131.111004} {\bibfield  {journal}
  {\bibinfo  {journal} {Phys. Rev. Lett.}\ }\textbf {\bibinfo {volume} {131}},\
  \bibinfo {pages} {111004} (\bibinfo {year} {2023})},\ \Eprint
  {https://arxiv.org/abs/2209.09917} {arXiv:2209.09917 [hep-ph]} \BibitemShut
  {NoStop}%
\bibitem [{\citenamefont {Dessert}\ \emph {et~al.}(2020)\citenamefont
  {Dessert}, \citenamefont {Foster},\ and\ \citenamefont
  {Safdi}}]{Dessert_2020}%
  \BibitemOpen
  \bibfield  {author} {\bibinfo {author} {\bibfnamefont {C.}~\bibnamefont
  {Dessert}}, \bibinfo {author} {\bibfnamefont {J.~W.}\ \bibnamefont
  {Foster}},\ and\ \bibinfo {author} {\bibfnamefont {B.~R.}\ \bibnamefont
  {Safdi}},\ }\bibfield  {journal} {\bibinfo  {journal} {Phys. Rev. Lett.}\
  }\textbf {\bibinfo {volume} {125}},\ \href
  {https://doi.org/10.1103/physrevlett.125.261102}
  {10.1103/physrevlett.125.261102} (\bibinfo {year} {2020})\BibitemShut
  {NoStop}%
\bibitem [{\citenamefont {Hoof}\ and\ \citenamefont
  {Schulz}(2023)}]{Hoof_2023}%
  \BibitemOpen
  \bibfield  {author} {\bibinfo {author} {\bibfnamefont {S.}~\bibnamefont
  {Hoof}}\ and\ \bibinfo {author} {\bibfnamefont {L.}~\bibnamefont {Schulz}},\
  }\href {https://doi.org/10.1088/1475-7516/2023/03/054} {\bibfield  {journal}
  {\bibinfo  {journal} {Journal of Cosmology and Astroparticle Physics}\
  }\textbf {\bibinfo {volume} {2023}}\bibinfo  {number} { (03)},\ \bibinfo
  {pages} {054}}\BibitemShut {NoStop}%
\bibitem [{\citenamefont {Manzari}\ \emph {et~al.}(2024)\citenamefont
  {Manzari}, \citenamefont {Park}, \citenamefont {Safdi},\ and\ \citenamefont
  {Savoray}}]{Manzari:2024jns}%
  \BibitemOpen
\bibfield  {number} {  }\bibfield  {author} {\bibinfo {author} {\bibfnamefont
  {C.~A.}\ \bibnamefont {Manzari}}, \bibinfo {author} {\bibfnamefont
  {Y.}~\bibnamefont {Park}}, \bibinfo {author} {\bibfnamefont {B.~R.}\
  \bibnamefont {Safdi}},\ and\ \bibinfo {author} {\bibfnamefont
  {I.}~\bibnamefont {Savoray}},\ }\href
  {https://doi.org/10.1103/PhysRevLett.133.211002} {\bibfield  {journal}
  {\bibinfo  {journal} {Phys. Rev. Lett.}\ }\textbf {\bibinfo {volume} {133}},\
  \bibinfo {pages} {211002} (\bibinfo {year} {2024})}\BibitemShut {NoStop}%
\bibitem [{\citenamefont {Foster}\ \emph {et~al.}(2022)\citenamefont {Foster}
  \emph {et~al.}}]{Foster:2022fxn}%
  \BibitemOpen
  \bibfield  {author} {\bibinfo {author} {\bibfnamefont {J.~W.}\ \bibnamefont
  {Foster}} \emph {et~al.},\ }\href
  {https://doi.org/10.1103/PhysRevLett.129.251102} {\bibfield  {journal}
  {\bibinfo  {journal} {Phys. Rev. Lett.}\ }\textbf {\bibinfo {volume} {129}},\
  \bibinfo {pages} {251102} (\bibinfo {year} {2022})},\ \Eprint
  {https://arxiv.org/abs/2202.08274} {arXiv:2202.08274 [astro-ph.CO]}
  \BibitemShut {NoStop}%
\bibitem [{\citenamefont {Battye}\ \emph {et~al.}(2023)\citenamefont {Battye}
  \emph {et~al.}}]{Battye:2023oac}%
  \BibitemOpen
  \bibfield  {author} {\bibinfo {author} {\bibfnamefont {R.~A.}\ \bibnamefont
  {Battye}} \emph {et~al.},\ }\href
  {https://doi.org/10.1103/PhysRevD.108.063001} {\bibfield  {journal} {\bibinfo
   {journal} {Phys. Rev. D}\ }\textbf {\bibinfo {volume} {108}},\ \bibinfo
  {pages} {063001} (\bibinfo {year} {2023})},\ \Eprint
  {https://arxiv.org/abs/2303.11792} {arXiv:2303.11792 [astro-ph.CO]}
  \BibitemShut {NoStop}%
\bibitem [{\citenamefont {Escudero}\ \emph {et~al.}(2024)\citenamefont
  {Escudero} \emph {et~al.}}]{Escudero:2023vgv}%
  \BibitemOpen
  \bibfield  {author} {\bibinfo {author} {\bibfnamefont {M.}~\bibnamefont
  {Escudero}} \emph {et~al.},\ }\href
  {https://doi.org/10.1103/PhysRevD.109.043018} {\bibfield  {journal} {\bibinfo
   {journal} {Phys. Rev. D}\ }\textbf {\bibinfo {volume} {109}},\ \bibinfo
  {pages} {043018} (\bibinfo {year} {2024})},\ \Eprint
  {https://arxiv.org/abs/2302.10206} {arXiv:2302.10206 [hep-ph]} \BibitemShut
  {NoStop}%
\bibitem [{\citenamefont {Fox}\ \emph {et~al.}(2023)\citenamefont {Fox},
  \citenamefont {Weiner},\ and\ \citenamefont {Xiao}}]{Fox:2023xgx}%
  \BibitemOpen
  \bibfield  {author} {\bibinfo {author} {\bibfnamefont {P.~J.}\ \bibnamefont
  {Fox}}, \bibinfo {author} {\bibfnamefont {N.}~\bibnamefont {Weiner}},\ and\
  \bibinfo {author} {\bibfnamefont {H.}~\bibnamefont {Xiao}},\ }\href
  {https://doi.org/10.1103/PhysRevD.108.095043} {\bibfield  {journal} {\bibinfo
   {journal} {Phys. Rev. D}\ }\textbf {\bibinfo {volume} {108}},\ \bibinfo
  {pages} {095043} (\bibinfo {year} {2023})},\ \Eprint
  {https://arxiv.org/abs/2302.00685} {arXiv:2302.00685 [hep-ph]} \BibitemShut
  {NoStop}%
\bibitem [{\citenamefont {Raffelt}(2008)}]{Raffelt2008}%
  \BibitemOpen
  \bibfield  {author} {\bibinfo {author} {\bibfnamefont {G.}~\bibnamefont
  {Raffelt}},\ }\bibinfo {title} {Astrophysical axion bounds},\ in\ \href
  {https://doi.org/10.1007/978-3-540-73518-2_3} {\emph {\bibinfo {booktitle}
  {Axions: Theory, Cosmology, and Experimental Searches}}}\ (\bibinfo
  {publisher} {Springer Berlin Heidelberg},\ \bibinfo {address} {Berlin,
  Heidelberg},\ \bibinfo {year} {2008})\ pp.\ \bibinfo {pages}
  {51--71}\BibitemShut {NoStop}%
\bibitem [{\citenamefont {Carenza}\ \emph {et~al.}(2024)\citenamefont {Carenza}
  \emph {et~al.}}]{Carenza:2024ehj}%
  \BibitemOpen
  \bibfield  {author} {\bibinfo {author} {\bibfnamefont {P.}~\bibnamefont
  {Carenza}} \emph {et~al.},\ }\href@noop {} {\bibinfo {title} {Axion
  astrophysics}} (\bibinfo {year} {2024}),\ \Eprint
  {https://arxiv.org/abs/2411.02492} {arXiv:2411.02492 [hep-ph]} \BibitemShut
  {NoStop}%
\bibitem [{\citenamefont {Raffelt}\ and\ \citenamefont
  {Stodolsky}(1988)}]{Raffelt:1987im}%
  \BibitemOpen
  \bibfield  {author} {\bibinfo {author} {\bibfnamefont {G.}~\bibnamefont
  {Raffelt}}\ and\ \bibinfo {author} {\bibfnamefont {L.}~\bibnamefont
  {Stodolsky}},\ }\href {https://doi.org/10.1103/PhysRevD.37.1237} {\bibfield
  {journal} {\bibinfo  {journal} {Phys. Rev. D}\ }\textbf {\bibinfo {volume}
  {37}},\ \bibinfo {pages} {1237} (\bibinfo {year} {1988})}\BibitemShut
  {NoStop}%
\bibitem [{\citenamefont {van Bibber}\ \emph {et~al.}(1989)\citenamefont {van
  Bibber}, \citenamefont {McIntyre}, \citenamefont {Morris},\ and\
  \citenamefont {Raffelt}}]{PhysRevD.39.2089}%
  \BibitemOpen
  \bibfield  {author} {\bibinfo {author} {\bibfnamefont {K.}~\bibnamefont {van
  Bibber}}, \bibinfo {author} {\bibfnamefont {P.~M.}\ \bibnamefont {McIntyre}},
  \bibinfo {author} {\bibfnamefont {D.~E.}\ \bibnamefont {Morris}},\ and\
  \bibinfo {author} {\bibfnamefont {G.~G.}\ \bibnamefont {Raffelt}},\ }\href
  {https://doi.org/10.1103/PhysRevD.39.2089} {\bibfield  {journal} {\bibinfo
  {journal} {Phys. Rev. D}\ }\textbf {\bibinfo {volume} {39}},\ \bibinfo
  {pages} {2089} (\bibinfo {year} {1989})}\BibitemShut {NoStop}%
\bibitem [{\citenamefont {Rempel}(2014)}]{Rempel_2014}%
  \BibitemOpen
  \bibfield  {author} {\bibinfo {author} {\bibfnamefont {M.}~\bibnamefont
  {Rempel}},\ }\href {https://doi.org/10.1088/0004-637x/789/2/132} {\bibfield
  {journal} {\bibinfo  {journal} {The Astrophysical Journal}\ }\textbf
  {\bibinfo {volume} {789}},\ \bibinfo {pages} {132} (\bibinfo {year}
  {2014})}\BibitemShut {NoStop}%
\bibitem [{\citenamefont {{V{\"o}gler}}\ \emph {et~al.}(2005)\citenamefont
  {{V{\"o}gler}} \emph {et~al.}}]{2005A&A...429..335V}%
  \BibitemOpen
  \bibfield  {author} {\bibinfo {author} {\bibfnamefont {A.}~\bibnamefont
  {{V{\"o}gler}}} \emph {et~al.},\ }\href
  {https://doi.org/10.1051/0004-6361:20041507} {\bibfield  {journal} {\bibinfo
  {journal} {Astronomy \& Astrophysics}\ }\textbf {\bibinfo {volume} {429}},\
  \bibinfo {pages} {335} (\bibinfo {year} {2005})}\BibitemShut {NoStop}%
\bibitem [{\citenamefont {{Rempel}}\ \emph {et~al.}(2009)\citenamefont
  {{Rempel}}, \citenamefont {{Sch{\"u}ssler}},\ and\ \citenamefont
  {{Kn{\"o}lker}}}]{2009ApJ...691..640R}%
  \BibitemOpen
  \bibfield  {author} {\bibinfo {author} {\bibfnamefont {M.}~\bibnamefont
  {{Rempel}}}, \bibinfo {author} {\bibfnamefont {M.}~\bibnamefont
  {{Sch{\"u}ssler}}},\ and\ \bibinfo {author} {\bibfnamefont {M.}~\bibnamefont
  {{Kn{\"o}lker}}},\ }\href {https://doi.org/10.1088/0004-637X/691/1/640}
  {\bibfield  {journal} {\bibinfo  {journal} {\apj}\ }\textbf {\bibinfo
  {volume} {691}},\ \bibinfo {pages} {640} (\bibinfo {year} {2009})},\ \Eprint
  {https://arxiv.org/abs/0808.3294} {arXiv:0808.3294 [astro-ph]} \BibitemShut
  {NoStop}%
\bibitem [{\citenamefont {{Trujillo Bueno}}\ \emph {et~al.}(2004)\citenamefont
  {{Trujillo Bueno}}, \citenamefont {{Shchukina}},\ and\ \citenamefont
  {{Asensio Ramos}}}]{2004Natur.430..326T}%
  \BibitemOpen
  \bibfield  {author} {\bibinfo {author} {\bibfnamefont {J.}~\bibnamefont
  {{Trujillo Bueno}}}, \bibinfo {author} {\bibfnamefont {N.}~\bibnamefont
  {{Shchukina}}},\ and\ \bibinfo {author} {\bibfnamefont {A.}~\bibnamefont
  {{Asensio Ramos}}},\ }\href {https://doi.org/10.1038/nature02669} {\bibfield
  {journal} {\bibinfo  {journal} {\nat}\ }\textbf {\bibinfo {volume} {430}},\
  \bibinfo {pages} {326} (\bibinfo {year} {2004})},\ \Eprint
  {https://arxiv.org/abs/arXiv:astro-ph/0409004} {arXiv:astro-ph/0409004}
  \BibitemShut {NoStop}%
\bibitem [{\citenamefont {{del Pino Aleman}}\ \emph {et~al.}(2018)\citenamefont
  {{del Pino Aleman}}, \citenamefont {{Trujillo Bueno}}, \citenamefont
  {Stepan},\ and\ \citenamefont {Shchukina}}]{PinoAleman2018}%
  \BibitemOpen
  \bibfield  {author} {\bibinfo {author} {\bibfnamefont {T.}~\bibnamefont {{del
  Pino Aleman}}}, \bibinfo {author} {\bibfnamefont {J.}~\bibnamefont {{Trujillo
  Bueno}}}, \bibinfo {author} {\bibfnamefont {J.}~\bibnamefont {Stepan}},\ and\
  \bibinfo {author} {\bibfnamefont {N.}~\bibnamefont {Shchukina}},\ }\href
  {https://doi.org/10.3847/1538-4357/aaceab} {\bibfield  {journal} {\bibinfo
  {journal} {The Astrophysical Journal}\ }\textbf {\bibinfo {volume} {863}},\
  \bibinfo {pages} {164} (\bibinfo {year} {2018})}\BibitemShut {NoStop}%
\bibitem [{\citenamefont {Mikić}\ \emph {et~al.}(2018)\citenamefont {Mikić}
  \emph {et~al.}}]{ps2017}%
  \BibitemOpen
  \bibfield  {author} {\bibinfo {author} {\bibfnamefont {Z.}~\bibnamefont
  {Mikić}} \emph {et~al.},\ }\href {https://doi.org/10.1038/s41550-018-0562-5}
  {\bibfield  {journal} {\bibinfo  {journal} {Nature Astronomy}\ }\textbf
  {\bibinfo {volume} {2}},\ \bibinfo {pages} {913–921} (\bibinfo {year}
  {2018})}\BibitemShut {NoStop}%
\bibitem [{sup()}]{supplemental_jra}%
  \BibitemOpen
  \href@noop {} {\bibinfo {title} {See supplemental material at}},\ \bibinfo
  {howpublished} {\url{(to_be_provided_by_editor)}},\ \bibinfo {note} {for
  further details, explanations, and figures related to the analysis methods
  and procedures used in this work, which includes Refs. [98-113] in addition
  to those cited in the main paper.}\BibitemShut {Stop}%
\bibitem [{\citenamefont {{Riley}}\ \emph {et~al.}(2006)\citenamefont
  {{Riley}}, \citenamefont {{Linker}}, \citenamefont {{Miki{\'c}}},
  \citenamefont {{Lionello}}, \citenamefont {{Ledvina}},\ and\ \citenamefont
  {{Luhmann}}}]{pffs6}%
  \BibitemOpen
  \bibfield  {author} {\bibinfo {author} {\bibfnamefont {P.}~\bibnamefont
  {{Riley}}}, \bibinfo {author} {\bibfnamefont {J.~A.}\ \bibnamefont
  {{Linker}}}, \bibinfo {author} {\bibfnamefont {Z.}~\bibnamefont
  {{Miki{\'c}}}}, \bibinfo {author} {\bibfnamefont {R.}~\bibnamefont
  {{Lionello}}}, \bibinfo {author} {\bibfnamefont {S.~A.}\ \bibnamefont
  {{Ledvina}}},\ and\ \bibinfo {author} {\bibfnamefont {J.~G.}\ \bibnamefont
  {{Luhmann}}},\ }\href {https://doi.org/10.1086/508565} {\bibfield  {journal}
  {\bibinfo  {journal} {\apj}\ }\textbf {\bibinfo {volume} {653}},\ \bibinfo
  {pages} {1510} (\bibinfo {year} {2006})}\BibitemShut {NoStop}%
\bibitem [{\citenamefont {{De La Luz}}\ \emph {et~al.}(2008)\citenamefont {{De
  La Luz}}, \citenamefont {{Lara}}, \citenamefont {{Mendoza}},\ and\
  \citenamefont {{Shimojo}}}]{2008GeofI..47..197D}%
  \BibitemOpen
  \bibfield  {author} {\bibinfo {author} {\bibfnamefont {V.}~\bibnamefont {{De
  La Luz}}}, \bibinfo {author} {\bibfnamefont {A.}~\bibnamefont {{Lara}}},
  \bibinfo {author} {\bibfnamefont {E.}~\bibnamefont {{Mendoza}}},\ and\
  \bibinfo {author} {\bibfnamefont {M.}~\bibnamefont {{Shimojo}}},\ }\href
  {https://ui.adsabs.harvard.edu/abs/2008GeofI..47..197D} {\bibfield  {journal}
  {\bibinfo  {journal} {Geofisica Internacional}\ }\textbf {\bibinfo {volume}
  {47}},\ \bibinfo {pages} {197} (\bibinfo {year} {2008})}\BibitemShut
  {NoStop}%
\bibitem [{\citenamefont {{Vernazza}}\ \emph {et~al.}(1981)\citenamefont
  {{Vernazza}}, \citenamefont {{Avrett}},\ and\ \citenamefont
  {{Loeser}}}]{1981ApJS...45..635V}%
  \BibitemOpen
  \bibfield  {author} {\bibinfo {author} {\bibfnamefont {J.~E.}\ \bibnamefont
  {{Vernazza}}}, \bibinfo {author} {\bibfnamefont {E.~H.}\ \bibnamefont
  {{Avrett}}},\ and\ \bibinfo {author} {\bibfnamefont {R.}~\bibnamefont
  {{Loeser}}},\ }\href {https://doi.org/10.1086/190731} {\bibfield  {journal}
  {\bibinfo  {journal} {\apjs}\ }\textbf {\bibinfo {volume} {45}},\ \bibinfo
  {pages} {635} (\bibinfo {year} {1981})}\BibitemShut {NoStop}%
\bibitem [{\citenamefont {{Gabriel}}(1976)}]{1976RSPTA.281..339G}%
  \BibitemOpen
  \bibfield  {author} {\bibinfo {author} {\bibfnamefont {A.~H.}\ \bibnamefont
  {{Gabriel}}},\ }\href {https://doi.org/10.1098/rsta.1976.0031} {\bibfield
  {journal} {\bibinfo  {journal} {Philosophical Transactions of the Royal
  Society of London Series A}\ }\textbf {\bibinfo {volume} {281}},\ \bibinfo
  {pages} {339} (\bibinfo {year} {1976})}\BibitemShut {NoStop}%
\bibitem [{\citenamefont {{Schmelz}}\ \emph {et~al.}(2012)\citenamefont
  {{Schmelz}} \emph {et~al.}}]{2012ApJ...755...33S}%
  \BibitemOpen
  \bibfield  {author} {\bibinfo {author} {\bibfnamefont {J.~T.}\ \bibnamefont
  {{Schmelz}}} \emph {et~al.},\ }\href
  {https://doi.org/10.1088/0004-637X/755/1/33} {\bibfield  {journal} {\bibinfo
  {journal} {\apj}\ }\textbf {\bibinfo {volume} {755}},\ \bibinfo {eid} {33}
  (\bibinfo {year} {2012})}\BibitemShut {NoStop}%
\bibitem [{\citenamefont {{Caffau}}\ \emph {et~al.}(2011)\citenamefont
  {{Caffau}} \emph {et~al.}}]{2011SoPh..268..255C}%
  \BibitemOpen
  \bibfield  {author} {\bibinfo {author} {\bibfnamefont {E.}~\bibnamefont
  {{Caffau}}} \emph {et~al.},\ }\href
  {https://doi.org/10.1007/s11207-010-9541-4} {\bibfield  {journal} {\bibinfo
  {journal} {Solar Physics}\ }\textbf {\bibinfo {volume} {268}},\ \bibinfo
  {pages} {255} (\bibinfo {year} {2011})},\ \Eprint
  {https://arxiv.org/abs/1003.1190} {arXiv:1003.1190 [astro-ph.SR]}
  \BibitemShut {NoStop}%
\bibitem [{\citenamefont {{Lodders}}\ \emph {et~al.}(2009)\citenamefont
  {{Lodders}}, \citenamefont {{Palme}},\ and\ \citenamefont
  {{Gail}}}]{2009LanB...4B..712L}%
  \BibitemOpen
  \bibfield  {author} {\bibinfo {author} {\bibfnamefont {K.}~\bibnamefont
  {{Lodders}}}, \bibinfo {author} {\bibfnamefont {H.}~\bibnamefont {{Palme}}},\
  and\ \bibinfo {author} {\bibfnamefont {H.-P.}\ \bibnamefont {{Gail}}},\
  }\href {https://doi.org/10.1007/978-3-540-88055-4_34} {\bibfield  {journal}
  {\bibinfo  {journal} {Landolt Börnstein}\ }\textbf {\bibinfo {volume}
  {4B}},\ \bibinfo {pages} {712} (\bibinfo {year} {2009})},\ \Eprint
  {https://arxiv.org/abs/0901.1149} {arXiv:0901.1149 [astro-ph.EP]}
  \BibitemShut {NoStop}%
\bibitem [{\citenamefont {{Dere}}\ \emph {et~al.}(1997)\citenamefont {{Dere}}
  \emph {et~al.}}]{1997A&AS..125..149D}%
  \BibitemOpen
  \bibfield  {author} {\bibinfo {author} {\bibfnamefont {K.~P.}\ \bibnamefont
  {{Dere}}} \emph {et~al.},\ }\href {https://doi.org/10.1051/aas:1997368}
  {\bibfield  {journal} {\bibinfo  {journal} {\aaps}\ }\textbf {\bibinfo
  {volume} {125}},\ \bibinfo {pages} {149} (\bibinfo {year}
  {1997})}\BibitemShut {NoStop}%
\bibitem [{\citenamefont {{Del Zanna}}\ \emph {et~al.}(2021)\citenamefont {{Del
  Zanna}} \emph {et~al.}}]{2021ApJ...909...38D}%
  \BibitemOpen
  \bibfield  {author} {\bibinfo {author} {\bibfnamefont {G.}~\bibnamefont {{Del
  Zanna}}} \emph {et~al.},\ }\href {https://doi.org/10.3847/1538-4357/abd8ce}
  {\bibfield  {journal} {\bibinfo  {journal} {\apj}\ }\textbf {\bibinfo
  {volume} {909}},\ \bibinfo {eid} {38} (\bibinfo {year} {2021})},\ \Eprint
  {https://arxiv.org/abs/2011.05211} {arXiv:2011.05211 [physics.atom-ph]}
  \BibitemShut {NoStop}%
\bibitem [{\citenamefont {{Berger}}\ \emph {et~al.}(2010)\citenamefont
  {{Berger}} \emph {et~al.}}]{xcom}%
  \BibitemOpen
  \bibfield  {author} {\bibinfo {author} {\bibfnamefont {M.~J.}\ \bibnamefont
  {{Berger}}} \emph {et~al.},\ }\href
  {https://doi.org/https://dx.doi.org/10.18434/T48G6X} {\bibinfo {title} {Nist
  standard reference database 8 (xgam)}} (\bibinfo {year} {2010})\BibitemShut
  {NoStop}%
\bibitem [{\citenamefont {Paterson}\ \emph {et~al.}(2024)\citenamefont
  {Paterson} \emph {et~al.}}]{sarah24}%
  \BibitemOpen
  \bibfield  {author} {\bibinfo {author} {\bibfnamefont {S.}~\bibnamefont
  {Paterson}} \emph {et~al.},\ }\href@noop {} {\bibfield  {journal} {\bibinfo
  {journal} {Monthly Notices of the Royal Astronomical Society}\ }\textbf
  {\bibinfo {volume} {528}},\ \bibinfo {pages} {6398} (\bibinfo {year}
  {2024})}\BibitemShut {NoStop}%
\bibitem [{\citenamefont {Andriamonje}\ \emph {et~al.}(2007)\citenamefont
  {Andriamonje} \emph {et~al.}}]{Andriamonje_2007}%
  \BibitemOpen
  \bibfield  {author} {\bibinfo {author} {\bibfnamefont {S.}~\bibnamefont
  {Andriamonje}} \emph {et~al.} (\bibinfo {collaboration} {CAST
  Collaboration}),\ }\href {https://doi.org/10.1088/1475-7516/2007/04/010}
  {\bibfield  {journal} {\bibinfo  {journal} {Journal of Cosmology and
  Astroparticle Physics}\ }\textbf {\bibinfo {volume} {2007}}\bibinfo  {number}
  { (04)},\ \bibinfo {pages} {010–010}}\BibitemShut {NoStop}%
\bibitem [{\citenamefont {Arnaud}\ and\ \citenamefont
  {et~al.}(1999)}]{arnaud1999xspec}%
  \BibitemOpen
\bibfield  {number} {  }\bibfield  {author} {\bibinfo {author} {\bibfnamefont
  {K.}~\bibnamefont {Arnaud}}\ and\ \bibinfo {author} {\bibnamefont {et~al.}},\
  }\href@noop {} {\bibfield  {journal} {\bibinfo  {journal} {Astrophysics
  Source Code Library}\ ,\ \bibinfo {pages} {ascl}} (\bibinfo {year}
  {1999})}\BibitemShut {NoStop}%
\bibitem [{\citenamefont {Brejnholt}\ \emph {et~al.}(2012)\citenamefont
  {Brejnholt} \emph {et~al.}}]{SPIE_NuSTAR_Effective}%
  \BibitemOpen
  \bibfield  {author} {\bibinfo {author} {\bibfnamefont {N.~F.}\ \bibnamefont
  {Brejnholt}} \emph {et~al.},\ }in\ \href {https://doi.org/10.1117/12.925631}
  {\emph {\bibinfo {booktitle} {Space Telescopes and Instrumentation 2012:
  Ultraviolet to Gamma Ray}}},\ Vol.\ \bibinfo {volume} {8443},\ \bibinfo
  {editor} {edited by\ \bibinfo {editor} {\bibfnamefont {T.}~\bibnamefont
  {Takahashi}}, \bibinfo {editor} {\bibfnamefont {S.~S.}\ \bibnamefont
  {Murray}},\ and\ \bibinfo {editor} {\bibfnamefont {J.-W.~A.}\ \bibnamefont
  {den Herder}}},\ \bibinfo {organization} {International Society for Optics
  and Photonics}\ (\bibinfo  {publisher} {SPIE},\ \bibinfo {year} {2012})\ p.\
  \bibinfo {pages} {84431Y}\BibitemShut {NoStop}%
\bibitem [{\citenamefont {Meyer}\ \emph {et~al.}(2013)\citenamefont {Meyer},
  \citenamefont {Horns},\ and\ \citenamefont {Raue}}]{transparency}%
  \BibitemOpen
  \bibfield  {author} {\bibinfo {author} {\bibfnamefont {M.}~\bibnamefont
  {Meyer}}, \bibinfo {author} {\bibfnamefont {D.}~\bibnamefont {Horns}},\ and\
  \bibinfo {author} {\bibfnamefont {M.}~\bibnamefont {Raue}},\ }\href
  {https://doi.org/10.1103/PhysRevD.87.035027} {\bibfield  {journal} {\bibinfo
  {journal} {Phys. Rev. D}\ }\textbf {\bibinfo {volume} {87}},\ \bibinfo
  {pages} {035027} (\bibinfo {year} {2013})}\BibitemShut {NoStop}%
\bibitem [{\citenamefont {Schlattl}\ \emph {et~al.}(1999)\citenamefont
  {Schlattl}, \citenamefont {Weiss},\ and\ \citenamefont
  {Raffelt}}]{SCHLATTL1999353}%
  \BibitemOpen
  \bibfield  {author} {\bibinfo {author} {\bibfnamefont {H.}~\bibnamefont
  {Schlattl}}, \bibinfo {author} {\bibfnamefont {A.}~\bibnamefont {Weiss}},\
  and\ \bibinfo {author} {\bibfnamefont {G.}~\bibnamefont {Raffelt}},\ }\href
  {https://doi.org/https://doi.org/10.1016/S0927-6505(98)00063-2} {\bibfield
  {journal} {\bibinfo  {journal} {Astroparticle Physics}\ }\textbf {\bibinfo
  {volume} {10}},\ \bibinfo {pages} {353} (\bibinfo {year} {1999})}\BibitemShut
  {NoStop}%
\bibitem [{\citenamefont {Vinyoles}\ \emph {et~al.}(2015)\citenamefont
  {Vinyoles} \emph {et~al.}}]{Vinyoles_2015}%
  \BibitemOpen
  \bibfield  {author} {\bibinfo {author} {\bibfnamefont {N.}~\bibnamefont
  {Vinyoles}} \emph {et~al.},\ }\href
  {https://doi.org/10.1088/1475-7516/2015/10/015} {\bibfield  {journal}
  {\bibinfo  {journal} {Journal of Cosmology and Astroparticle Physics}\
  }\textbf {\bibinfo {volume} {2015}}\bibinfo  {number} { (10)},\ \bibinfo
  {pages} {015}}\BibitemShut {NoStop}%
\bibitem [{\citenamefont {Hoof}\ \emph {et~al.}(2021)\citenamefont {Hoof},
  \citenamefont {Jaeckel},\ and\ \citenamefont {Thormaehlen}}]{Hoof:2021mld}%
  \BibitemOpen
\bibfield  {number} {  }\bibfield  {author} {\bibinfo {author} {\bibfnamefont
  {S.}~\bibnamefont {Hoof}}, \bibinfo {author} {\bibfnamefont {J.}~\bibnamefont
  {Jaeckel}},\ and\ \bibinfo {author} {\bibfnamefont {L.~J.}\ \bibnamefont
  {Thormaehlen}},\ }\href {https://doi.org/10.1088/1475-7516/2021/09/006}
  {\bibfield  {journal} {\bibinfo  {journal} {Journal of Cosmology and
  Astroparticle Physics}\ }\textbf {\bibinfo {volume} {2021}}\bibinfo  {number}
  { (09)},\ \bibinfo {pages} {006}}\BibitemShut {NoStop}%
\bibitem [{\citenamefont {Todarello}(2025)}]{elisa_git}%
  \BibitemOpen
\bibfield  {number} {  }\bibfield  {author} {\bibinfo {author} {\bibfnamefont
  {E.}~\bibnamefont {Todarello}},\ }\href@noop {} {\bibinfo {title}
  {Nustar-helioscop code repository}},\ \bibinfo {howpublished}
  {\url{https://github.com/elisabm99/nustar-helioscope}} (\bibinfo {year}
  {2025}),\ \bibinfo {note} {accessed: 2025-10-09}\BibitemShut {NoStop}%
\bibitem [{\citenamefont {Bahcall}\ \emph {et~al.}(1982)\citenamefont {Bahcall}
  \emph {et~al.}}]{bahcall1982standard}%
  \BibitemOpen
  \bibfield  {author} {\bibinfo {author} {\bibfnamefont {J.~N.}\ \bibnamefont
  {Bahcall}} \emph {et~al.},\ }\href
  {https://doi.org/10.1103/RevModPhys.54.767} {\bibfield  {journal} {\bibinfo
  {journal} {Reviews of Modern Physics}\ }\textbf {\bibinfo {volume} {54}},\
  \bibinfo {pages} {767} (\bibinfo {year} {1982})}\BibitemShut {NoStop}%
\bibitem [{\citenamefont {Arik}\ \emph {et~al.}(2009)\citenamefont {Arik} \emph
  {et~al.}}]{cast_jcap2009}%
  \BibitemOpen
  \bibfield  {author} {\bibinfo {author} {\bibfnamefont {E.}~\bibnamefont
  {Arik}} \emph {et~al.} (\bibinfo {collaboration} {CAST Collaboration}),\
  }\href {https://doi.org/10.1088/1475-7516/2009/02/008} {\bibfield  {journal}
  {\bibinfo  {journal} {Journal of Cosmology and Astroparticle Physics}\
  }\textbf {\bibinfo {volume} {2009}}\bibinfo  {number} { (02)},\ \bibinfo
  {pages} {008}}\BibitemShut {NoStop}%
\bibitem [{\citenamefont {{Lites}}\ \emph {et~al.}(2008)\citenamefont {{Lites}}
  \emph {et~al.}}]{2008ApJ...672.1237L}%
  \BibitemOpen
\bibfield  {number} {  }\bibfield  {author} {\bibinfo {author} {\bibfnamefont
  {B.~W.}\ \bibnamefont {{Lites}}} \emph {et~al.},\ }\href
  {https://doi.org/10.1086/522922} {\bibfield  {journal} {\bibinfo  {journal}
  {\apj}\ }\textbf {\bibinfo {volume} {672}},\ \bibinfo {pages} {1237}
  (\bibinfo {year} {2008})}\BibitemShut {NoStop}%
\bibitem [{\citenamefont {{Kariyappa}}\ \emph {et~al.}(2011)\citenamefont
  {{Kariyappa}} \emph {et~al.}}]{2011A&A...526A..78K}%
  \BibitemOpen
  \bibfield  {author} {\bibinfo {author} {\bibfnamefont {R.}~\bibnamefont
  {{Kariyappa}}} \emph {et~al.},\ }\href
  {https://doi.org/10.1051/0004-6361/201014878} {\bibfield  {journal} {\bibinfo
   {journal} {Astronomy \& Astrophysics}\ }\textbf {\bibinfo {volume} {526}},\
  \bibinfo {eid} {A78} (\bibinfo {year} {2011})}\BibitemShut {NoStop}%
\bibitem [{iai()}]{iain_github}%
  \BibitemOpen
  \href {http://ianan.github.io/nsovr} {\emph {\bibinfo {title} {NuSTAR
  quickplots}}}\BibitemShut {NoStop}%
\bibitem [{\citenamefont {Grefenstette}\ \emph {et~al.}(2016)\citenamefont
  {Grefenstette} \emph {et~al.}}]{Grefenstette_2016}%
  \BibitemOpen
  \bibfield  {author} {\bibinfo {author} {\bibfnamefont {B.~W.}\ \bibnamefont
  {Grefenstette}} \emph {et~al.},\ }\href
  {https://doi.org/10.3847/0004-637X/826/1/20} {\bibfield  {journal} {\bibinfo
  {journal} {The Astrophysical Journal}\ }\textbf {\bibinfo {volume} {826}},\
  \bibinfo {pages} {20} (\bibinfo {year} {2016})}\BibitemShut {NoStop}%
\bibitem [{\citenamefont {Dessler}(1959)}]{Dessler1959}%
  \BibitemOpen
  \bibfield  {author} {\bibinfo {author} {\bibfnamefont {A.~J.}\ \bibnamefont
  {Dessler}},\ }\href {https://doi.org/https://doi.org/10.1029/JZ064i007p00713}
  {\bibfield  {journal} {\bibinfo  {journal} {Journal of Geophysical Research
  (1896-1977)}\ }\textbf {\bibinfo {volume} {64}},\ \bibinfo {pages} {713}
  (\bibinfo {year} {1959})}\BibitemShut {NoStop}%
\bibitem [{\citenamefont {Perri}\ \emph {et~al.}(2021)\citenamefont {Perri}
  \emph {et~al.}}]{Perri2020}%
  \BibitemOpen
  \bibfield  {author} {\bibinfo {author} {\bibfnamefont {M.}~\bibnamefont
  {Perri}} \emph {et~al.},\ }\href
  {https://heasarc.gsfc.nasa.gov/docs/nustar/analysis/nustar_swguide.pdf}
  {\emph {\bibinfo {title} {The NuSTAR Data Analysis Software Guide. Version
  1.9.7}}}\ (\bibinfo {year} {2021})\BibitemShut {NoStop}%
\bibitem [{\citenamefont {Wik}\ \emph {et~al.}(2014)\citenamefont {Wik} \emph
  {et~al.}}]{Wik_2014}%
  \BibitemOpen
  \bibfield  {author} {\bibinfo {author} {\bibfnamefont {D.~R.}\ \bibnamefont
  {Wik}} \emph {et~al.},\ }\href {https://doi.org/10.1088/0004-637X/792/1/48}
  {\bibfield  {journal} {\bibinfo  {journal} {The Astrophysical Journal}\
  }\textbf {\bibinfo {volume} {792}},\ \bibinfo {pages} {48} (\bibinfo {year}
  {2014})}\BibitemShut {NoStop}%
\bibitem [{\citenamefont {Rossland}\ \emph {et~al.}(2023)\citenamefont
  {Rossland} \emph {et~al.}}]{Rossland_2023}%
  \BibitemOpen
  \bibfield  {author} {\bibinfo {author} {\bibfnamefont {S.}~\bibnamefont
  {Rossland}} \emph {et~al.},\ }\href
  {https://doi.org/10.3847/1538-3881/acd0ae} {\bibfield  {journal} {\bibinfo
  {journal} {The Astronomical Journal}\ }\textbf {\bibinfo {volume} {166}},\
  \bibinfo {pages} {20} (\bibinfo {year} {2023})}\BibitemShut {NoStop}%
\bibitem [{\citenamefont {Arik}\ \emph {et~al.}(2011)\citenamefont {Arik} \emph
  {et~al.}}]{CAST_PRL2011}%
  \BibitemOpen
  \bibfield  {author} {\bibinfo {author} {\bibfnamefont {M.}~\bibnamefont
  {Arik}} \emph {et~al.} (\bibinfo {collaboration} {CAST Collaboration}),\
  }\href {https://doi.org/10.1103/PhysRevLett.107.261302} {\bibfield  {journal}
  {\bibinfo  {journal} {Phys. Rev. Lett.}\ }\textbf {\bibinfo {volume} {107}},\
  \bibinfo {pages} {261302} (\bibinfo {year} {2011})}\BibitemShut {NoStop}%
\bibitem [{\citenamefont {Arik}\ \emph {et~al.}(2014)\citenamefont {Arik} \emph
  {et~al.}}]{PhysRevLett.112.091302}%
  \BibitemOpen
  \bibfield  {author} {\bibinfo {author} {\bibfnamefont {M.}~\bibnamefont
  {Arik}} \emph {et~al.} (\bibinfo {collaboration} {CAST Collaboration}),\
  }\href {https://doi.org/10.1103/PhysRevLett.112.091302} {\bibfield  {journal}
  {\bibinfo  {journal} {Phys. Rev. Lett.}\ }\textbf {\bibinfo {volume} {112}},\
  \bibinfo {pages} {091302} (\bibinfo {year} {2014})}\BibitemShut {NoStop}%
\bibitem [{\citenamefont {Anastassopoulos}\ \emph {et~al.}(2017)\citenamefont
  {Anastassopoulos} \emph {et~al.}}]{CAST_nature}%
  \BibitemOpen
  \bibfield  {author} {\bibinfo {author} {\bibfnamefont {V.}~\bibnamefont
  {Anastassopoulos}} \emph {et~al.} (\bibinfo {collaboration} {CAST
  Collaboration}),\ }\href {https://doi.org/10.1038/nphys4109} {\bibfield
  {journal} {\bibinfo  {journal} {Nature Physics}\ }\textbf {\bibinfo {volume}
  {13}},\ \bibinfo {pages} {584} (\bibinfo {year} {2017})}\BibitemShut
  {NoStop}%
\bibitem [{\citenamefont {Zioutas}\ \emph {et~al.}(2005)\citenamefont {Zioutas}
  \emph {et~al.}}]{CAST_PRL04}%
  \BibitemOpen
  \bibfield  {author} {\bibinfo {author} {\bibfnamefont {K.}~\bibnamefont
  {Zioutas}} \emph {et~al.} (\bibinfo {collaboration} {CAST Collaboration}),\
  }\href {https://doi.org/10.1103/PhysRevLett.94.121301} {\bibfield  {journal}
  {\bibinfo  {journal} {Phys. Rev. Lett.}\ }\textbf {\bibinfo {volume} {94}},\
  \bibinfo {pages} {121301} (\bibinfo {year} {2005})}\BibitemShut {NoStop}%
\bibitem [{\citenamefont {Ayala}\ \emph {et~al.}(2014)\citenamefont {Ayala}
  \emph {et~al.}}]{Ayala_2014}%
  \BibitemOpen
  \bibfield  {author} {\bibinfo {author} {\bibfnamefont {A.}~\bibnamefont
  {Ayala}} \emph {et~al.},\ }\bibfield  {journal} {\bibinfo  {journal} {Phys.
  Rev. Lett.}\ }\textbf {\bibinfo {volume} {113}},\ \href
  {https://doi.org/10.1103/physrevlett.113.191302}
  {10.1103/physrevlett.113.191302} (\bibinfo {year} {2014})\BibitemShut
  {NoStop}%
\bibitem [{\citenamefont {Dolan}\ \emph {et~al.}(2022)\citenamefont {Dolan},
  \citenamefont {Hiskens},\ and\ \citenamefont {Volkas}}]{Dolan_2022}%
  \BibitemOpen
  \bibfield  {author} {\bibinfo {author} {\bibfnamefont {M.~J.}\ \bibnamefont
  {Dolan}}, \bibinfo {author} {\bibfnamefont {F.~J.}\ \bibnamefont {Hiskens}},\
  and\ \bibinfo {author} {\bibfnamefont {R.~R.}\ \bibnamefont {Volkas}},\
  }\href {https://doi.org/10.1088/1475-7516/2022/10/096} {\bibfield  {journal}
  {\bibinfo  {journal} {Journal of Cosmology and Astroparticle Physics}\
  }\textbf {\bibinfo {volume} {2022}}\bibinfo  {number} { (10)},\ \bibinfo
  {pages} {096}}\BibitemShut {NoStop}%
\end{thebibliography}%
